\documentclass[fleqn,usenatbib,useAMS]{mnras}

\usepackage{graphicx,amsmath,amssymb,multicol,bm,pdflscape, hyperref, soul, graphicx,url}	
\usepackage[T1]{fontenc}
\usepackage[normalem]{ulem}
\usepackage{newtxtext,newtxmath}

\title[Noiseless Astronomy]{Learning to Denoise Astronomical Images  with U-nets\\}

\author[A. Vojtekova]{Antonia Vojtekova$^{1,2}$\thanks{Contact e-mail: \href{mailto: vojtekova.antonia@gmail.com}{vojtekova.antonia@gmail.com}}, Maggie Lieu$^{2,3}$, 
Ivan Valtchanov$^4$, 
Bruno Altieri$^2$, 
Lyndsay Old$^2$,
\newauthor 
Qifeng Chen$^5$, Filip Hroch$^1$
\\
$^1$ Department of Theoretical Physics and Astrophysics, Masaryk University, Brno, Czech Republic\\
$^{2}$ European Space Agency, ESAC, Camino Bajo del Castillo, 28692, Villanueva de la Ca$\tilde{n}$ada, Madrid, Spain \\
$^{3}$ School of Physics $\&$ Astronomy, University of Nottingham, Nottingham, NG7 2RD \\
$^{4}$Telespazio Vega UK for ESA, European Space Astronomy Centre, Operations Department, 28691 Villanueva de la Ca\~nada, Spain\\
$^5$ Department of Computer Science and Engineering, The Hong Kong University of Science and Technology, Hong Kong\\
}

\date{Last updated 2015 May 22; in original form 2013 September 5}

\pubyear{2019}

\begin{document}
\label{firstpage}
\pagerange{\pageref{firstpage}--\pageref{lastpage}}
\maketitle

\begin{abstract}
Astronomical images are essential for exploring and understanding the universe. Optical telescopes capable of deep observations, such as the Hubble Space Telescope, are heavily oversubscribed in the Astronomical Community. Images also often contain additive noise, which makes de-noising a mandatory step in post-processing the data before further data analysis.  In order to maximise the efficiency and information gain in the post-processing of astronomical imaging, we turn to machine learning. We propose \texttt{Astro U-net}, a convolutional neural network for image de-noising and enhancement. For a proof-of-concept, we use Hubble space telescope images from WFC3 instrument UVIS with F555W and F606W filters. Our network is able to produce images with noise characteristics as if they are obtained with twice the exposure time, and with minimum bias or information loss. From these images, we are able to recover $95.9\%$ of stars with an average flux error of $2.26\%$. Furthermore the images have, on average, $1.63$ times higher signal-to-noise ratio than the input noisy images, equivalent to the stacking of at least $3$ input images,  which means a significant reduction in the telescope time needed for future astronomical imaging campaigns.

\end{abstract}



\section{Introduction}
In astronomy, besides the challenges of building and developing state-of-the-art telescopes and cameras, another challenge for optimising the information gain from images is the image processing pipeline. Noise affects our ability to extract the signals we are interested in, and given a fixed amount of telescope time, adds limitations to our science. Instrumental noise, such as the dark current and readout noise can be reduced by cooling the camera and by subtracting the dark and flat images. It is, however, non-trivial to completely subtract the noise from the detected signal. The signal-to-noise ratio (SNR) is an important metric for astronomical observations which represents the amount information in the data in comparison to the noise. 

It is possible to obtain better imaging by increasing the exposure time to obtain a higher signal, however, this is at the expense of higher noise. It is also possible to stack multiple short exposure images to increase SNR. This option increases the depth of image, reduces the noise and decreases the number of cosmic rays, however, this approach requires long observations and long processing times to select good images, align and combine them \citep{Zackay20171, Zackay20172}.  

Further image enhancement to improve noise levels can be achieved by applying linear or non-linear filters. The Gaussian filter \citep[e.g. Gaussian][]{Seddik2012,Pourebrahimi2009} is a commonly used low-pass linear filter that reduces high frequency signals, and the median filter is a non-linear filter where the central pixel is replaced by the median of its neighboring pixels within the kernel \citep{Zhu2012,Charmouti2017}. In comparison with Gaussian filter, the median filter is better at preserving edges. There exists many other filters used for noise reduction, the choice of which is a fundamental problem not just in astronomy but for entire field of computer vision. 

\begin{figure*}
\includegraphics[width=\textwidth]{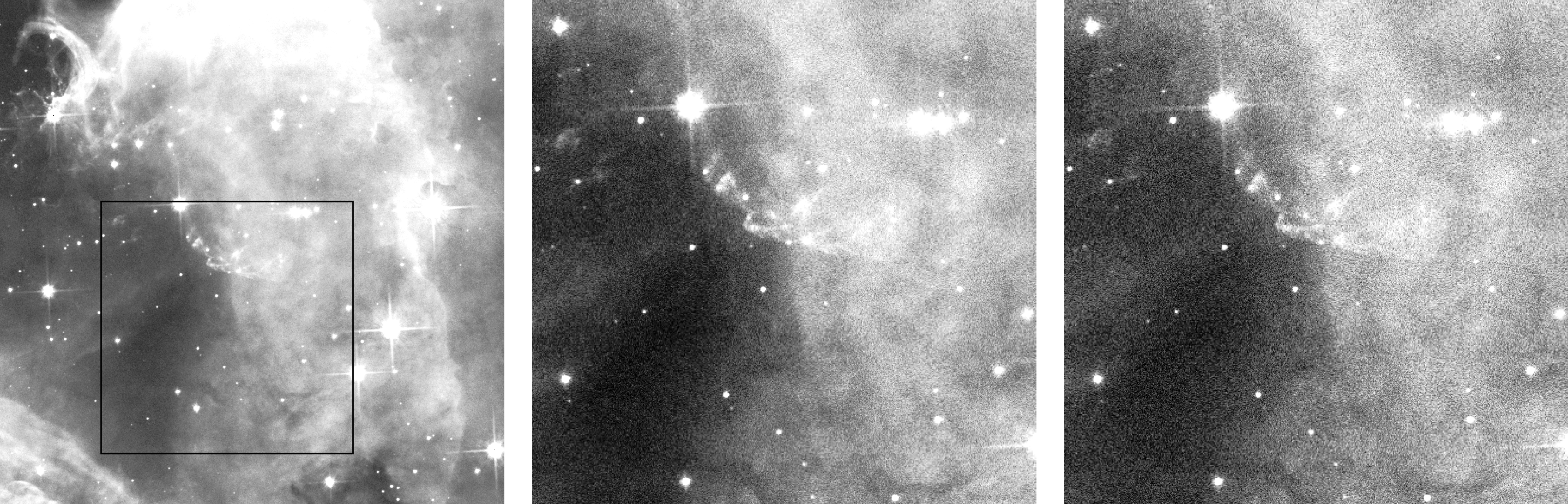}
\caption{An example validation image of nebula NGC 3603. The images have been scaled for visualisation purposes to the same scale, with the ground truth image (left), 1/2 exposure image (middle) and 1/5 exposure image (right).}
\label{fig:training_example}
\end{figure*}

Over the past decade, machine learning approaches including sparse, de-noising auto-encoders \citep{Agostinelli2013},  de-noising using dictionary learning \citep{Beckouche2013} and neural networks \citep{Schawinski2017}, have been introduced to improve imaging enhancement. More recently, convolutional neural networks (CNNs) have been widely used for a variety of image processing tasks such as de-noising, style transfer \citep{2017Chen}, super-resolution \citep{2019Zhang}, segmentation \citep{Carreira2012}, etc. For image-to-image mapping, networks usually consist of convolutional and deconvolutional layers. For this purpose, one can use fully-convolutional 
networks \citep{2014Long}, generative adversarial networks \citep[GAN,][]{Goodfellow2014}, multi-scale context aggregation networks \citep[CAN,][]{CAN} or U-net \citep{Ronneberger2015}, etc. 
With many upcoming, large astronomical surveys, it is now critical to find less time consuming approaches to analyse data. Neural networks have already been employed for many different tasks in astronomy, including cosmic web simulations \citep{Rodriguez2018}, to retrieve exoplanetary atmospheres \citep{Zingales2018} and image reconstruction \citep{Flamary2016, Baron2019}. In particular, \cite{Schawinski2017} use a GAN to recover features of an artificially degraded image, and \cite{Chen2018} present U-net for image enhancement and de-noising. Inspired by their work, we propose \texttt{Astro U-net}, the main goal of which is to reduce observation times and reduce noise in astronomical images
intended for scientific analysis.  

The layout of this paper is as follows, in \autoref{Data} we introduce the data set use for training and evaluation of the network and describe the process for obtaining the input images in our data set. In \autoref{Method} we introduce neural networks and we describe the architecture of Astro U-net. The results of the top-performing networks (\textit{Network 1} and \textit{2}) are presented in \autoref{Results}.  In addition to the top-performing networks we also discuss different hyper-parameters of the network, and discuss their influence on the results. In the \autoref{Discussion and conclusion} we discuss the results and conclusions.

\section{Data} \label{Data}
Our data set contains 200 images from the Hubble Space Telescope (HST) archives (\autoref{fig:training_example}) that are divided into training, evaluation and validation data sets of 160/20/20 images respectively. The data set is manually selected -- we went through more than a thousand images to create the final data set which includes astronomical objects of various scales. The images are captured by the UVIS (UV/Visible channel) detector on the Wide Field Camera 3 (WFC3), which is a $ \sim 4000 \times 4000$ pixel detector of two CCDs ($2051 \times 4096$ each), and a $31$ pixel gap between them. From a broad variety of filters, we select two wide filters -- F555W and F606W that correspond to 530.8 and 588.7 nm pivot wavelengths. We choose two wide filters, because we want to cover a wide range of the visible spectrum. The image sizes are significantly larger than the network input and therefore overfitting is not a large concern, despite the small number of images.

Henceforth we refer to these images as the \textit{real data}. They are used as the \textit{ground truth} when training the network. For the input to the network we use synthetic data that are generated based on the real data but with additional noise and shorter exposure times (see \autoref{Simulations}). The input to our network are \texttt{FITS} images in $electrons/s$ unless stated otherwise, and therefore we don't apply any other normalization. The evaluation is performed on images in $electrons$.  The images presented in this paper are created with Zscale and linear stretch adapted for visualisation purposes only.

\subsection{Simulations}  \label{Simulations}
The observed signal in astronomical images are degraded by various types of noise. To create synthetic data for the network input we consider photon shot noise, dark noise, and read-out noise. The number of photons detected from distant sources have an inherent statistical variation and the noise associated with it is: 
\begin{eqnarray}
N_\mathrm{photon} = {\rm Pois}(S),
\end{eqnarray}
where $S$ is total signal captured by the camera and ${\rm Pois}(X)$ is the Poisson distribution of $X$. Dark noise arises from thermally excited electrons of the detector. It is strongly dependent on the temperature of the CCD and is independent from photons falling on the detector and therefore this noise persists even when the camera is in complete darkness. The noise is also Poisson distributed but we use a Gaussian approximation calculated from the dark current ($\mathrm{DK}$):
\begin{eqnarray}
N_\mathrm{dark} = \mathcal{N}(0,\sqrt{\mathrm{DK} \cdot t}),
\end{eqnarray}
where $\textit{t}$ is the exposure time of the image. 

Lastly, the read-out noise ($\mathrm{RON}$) is a uniform noise across all pixels caused by the electronics of the CCD. 
We use the read-out noise and dark current values from the WFC3 Instrument Handbook \footnote{\url{https://hst-docs.stsci.edu/display/WFC3IHB/WFC3+Instrument+Handbook}}.
Before we add the noise, we create synthetic short-exposure images by dividing the real images $I_\mathrm{long}$ by the exposure time ratio, $r$:
\begin{eqnarray}
I_\mathrm{short} = I_\mathrm{long}/r.
\label{eq:exp_time_ratio}
\end{eqnarray}
The exposure time ratio is the ratio between the exposure time of the real image and the exposure time of the shorter simulated image. Examples of different ratios are shown in \autoref{fig:training_example}. Then the final synthetic data that are used in the network are generated as follows

\begin{eqnarray}
I_\mathrm{input} &=&  N_\mathrm{photon} + N_\mathrm{dark} + N_\mathrm{RON}  \nonumber \\
 &=& {\rm Pois}(I_\mathrm{short}) + \mathcal{N}(0,\sqrt{\mathrm{DK}\cdot t}) + \mathcal{N}(0, \mathrm{RON}), \nonumber \\
\end{eqnarray}
where $I_\mathrm{input}$ is the noisy input to the network and $I_\mathrm{long}$ is the corresponding high SNR output of the network. Note however, due to the large size of the images, we only feed the network random crops of the full images with size $256 \times 256$, in each iteration. 
 
\section{Method} \label{Method}
To denoise and enhance the astronomical images, we use convolutional neural networks. Whilst many different CNN architectures exist, we opt for the U-net architecture \citep{Ronneberger2015}.

\subsection{Convolutional neural networks} \label{sec:CNN}
Any image can be represented as 3-D matrix of size height $\times$ width $\times$ depth. To process grid-like data with a neural network, we can employ a CNN (\cite{lecun1990}, \cite{ciresan2012}, \cite{Goodfellow-et-al-2016}).

\begin{figure}
    \includegraphics[width=\columnwidth]{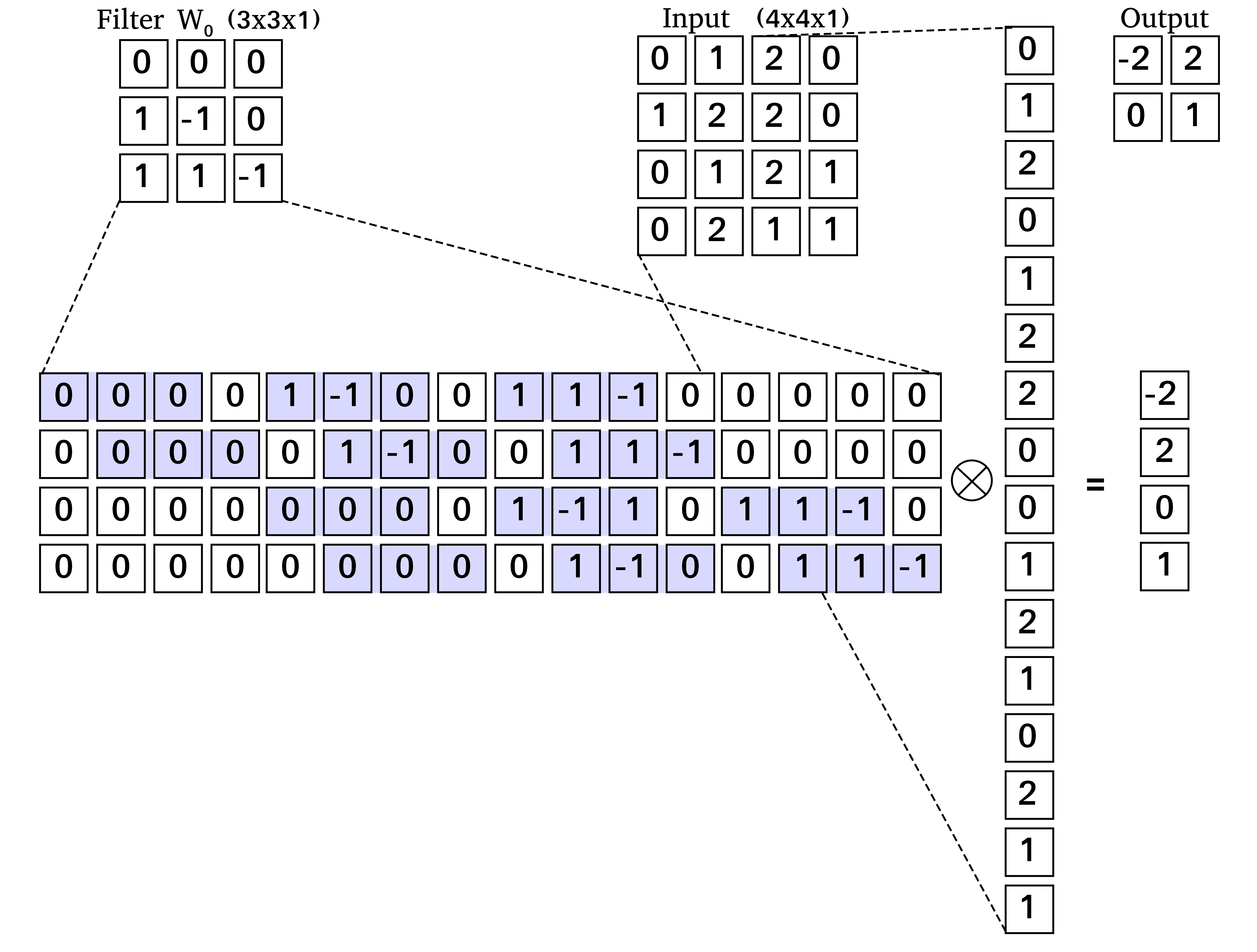}
    \caption{Example of the convolutional layer in the CNN. From the left: convolutional filter with size $3 \times 3 \times 1$, input feature map $4 \times 4 \times 1$ without zero-padding and output feature map with size $2 \times 2 \times 1$.  It is possible to represent the convolution operation as matrix multiplication. First, we have to rearrange filter into matrix, flatten the input and then the matrix multiplication can be done. Finally, the output has to be reshaped into the required shape.}
    \label{fig:cnn}
\end{figure}

A CNN consists mainly of convolutional layers that take an input feature map and convolves it with a set of filters to produce an output feature map:
\begin{eqnarray}
   O = (I * F)(i,j) = \sum_{m}\sum_{n}I_{m,n}F_{i-m, j-n},
\end{eqnarray}
where $I$ is the input for the layer, $F$ is the convolutional filter, and $O$ is the layer output (\autoref{fig:cnn}). Each filter contains trainable parameters (weights) that are updated during the training of network, in order to learn different types of features present in the data. The size of the output is determined by the filter size, stride and zero-padding. The stride denotes how much the filter slides on the input feature map between convolutions. Zero-padding is a margin of zeroes around the image border that control the spatial size of the output. The convolution layer is usually followed by a non-linear activation function. 

The spacial size (width and height) of the layers can be reduced by the convolutional layer, but more often the pooling layer (\cite{Boureau2010},\cite{Wu2015}, \cite{Scherer2010}) is used. This also reduces the number of computations performed by the network. Pooling is applied to every feature map in the layer independently using some pre-defined operation e.g. maximum, average.  In max-pooling, the maximum element is returned from an image patch overlapping with the filter. The filter is then moved across by the pre-defined stride and the process is repeated.

\begin{figure}
    \centering
    \includegraphics[width=0.56\columnwidth]{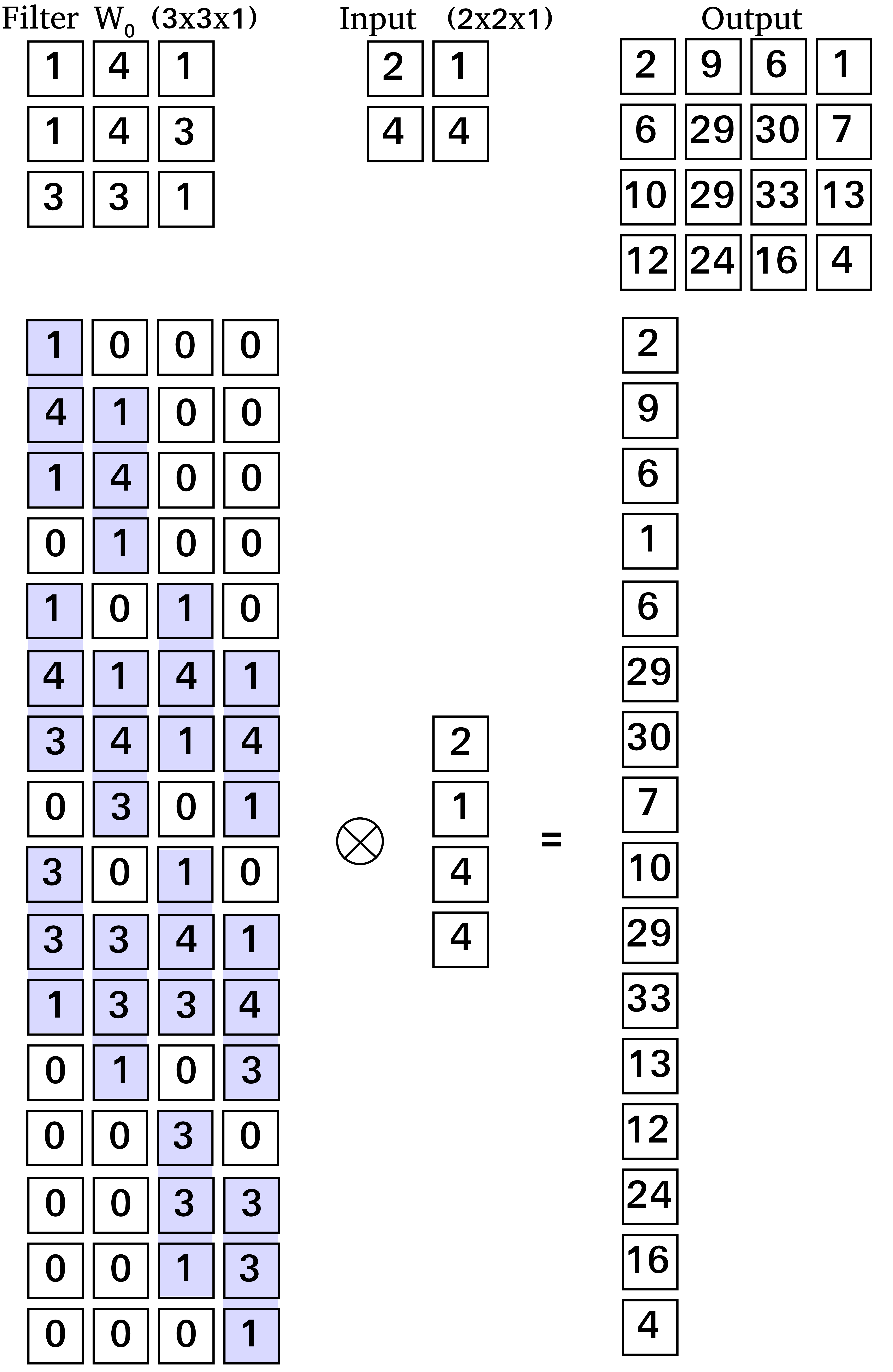}
    \caption{An example of the transposed convolution layer. The main purpose of the layer is to up-sample the input.  Here one can see the convolution operation as a matrix multiplication of the unrolled \textit{transpose} filter with the vectorised input. }
    \label{fig:trans}
\end{figure}
Another type of layer which can be used in the CNN is an up-sampling layer \citep{Dumoulin2016}. In up-sampling, the spacial size of the input is increased. This can be done using an up-sampling technique such as nearest neighbour or bi-linear interpolation algorithms, or by using transposed convolution. Interpolation methods use known data to estimate the unknown -- missing data \citep{Amanatiadis2009}. Nearest neighbour interpolation is a very fast and easy to implement algorithm - it takes the value of the nearest pixel and duplicates it onto the pixel with the unknown value. Bi-linear interpolation is a linear interpolation in two directions -- across the image height and width. It uses a weighted average of 4 surrounding pixels to approximate the value of the unknown pixel value, and the weight depends on the distance of known pixels to missing pixels. 

In comparison to the interpolation methods, transposed convolution has learnable parameters. Like the convolution layer, it requires the specification of the number of filters, their size and stride. All values in the filter are trainable parameters which are updated during the training of the network. The transposed convolution is named as it is, because if you write the convolution matrix as an unrolled matrix of the kernel multiplied by the vectorised input, then the transposed convolution is the equivalent but with the transpose of the unrolled kernel matrix multiplied by the vectorised input (\autoref{fig:trans}). We note that in transposed convolution, uneven overlap of areas on the output feature map can lead to checkerboard artifacts, so it is important to find a reasonable ratio between filter size and stride to avoid this problem.
 
Before training the network we have to specify the loss function, the type of optimizer, and choose the data set. In supervised learning \citep{Goodfellow-et-al-2016}, the data set consists of labelled training data e.g. an image of a galaxy might have the label of the type of galaxy. 
All weights are initiated randomly with small non-zero numbers. During training we want to find a set of weights which minimizes the loss function. We discuss the choice of loss function in \autoref{loss}. Note that the loss function is dependent on all of the weights in the network.  For this purpose we need to choose one of the optimizing algorithms (\cite{ribbinsMonro1951}, \cite{Bottou2018OptimizationMF}) . The most basic of these algorithms is gradient descent \citep{Dogo2018}, where the gradient is defined as the partial derivative of every input variable. It can be imagined as a vector pointing to the greatest increase of the function. To minimize the loss function, we take the opposite direction of the gradient of the loss function with respect to its weights. The gradient is calculated with respect to every weight and backpropagated \citep{Lecun2015} through the network:
\begin{eqnarray}\label{eqn:backprop}
    \textbf{w}^{t} = \textbf{w}^{t-1} - \alpha \nabla L(\textbf{w}) 
\end{eqnarray}
where $\textbf{w}$ denotes the weight matrix at the current $t$ and previous $t-1$ epoch and $L$ is the loss function and $\alpha$ is the learning rate. The learning rate is a hyper-parameter which determines the step with which the weights will change.
The training of the neural network is an iterative process and can be summarized as:
\begin{itemize}
    \item Forward pass --  training data goes into the network and the output is calculated.
    \item Loss function -- the output of the network is compared with the real data label and the loss is computed.
    \item Backpropagation -- the weights are adjusted according to \autoref{eqn:backprop}. 
\end{itemize}
This is iterated over for a number of epochs.

 \begin{figure*}
 \includegraphics[width=0.6\textwidth]{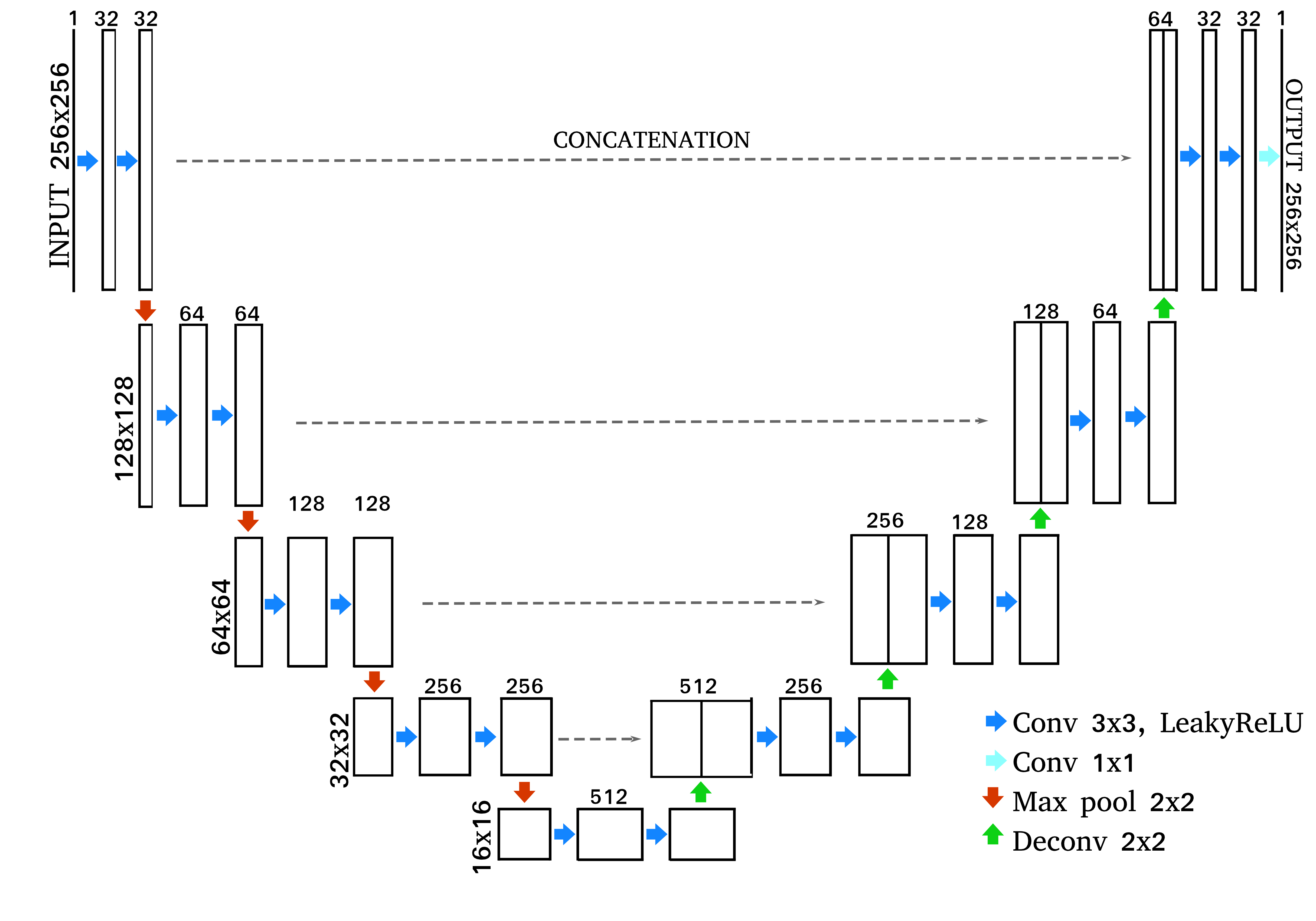}
 \caption{Architecture of the fully-convolutional network -- U-net. The input and output size is $256 \times 256$. The different colored arrows represent operations between layers. The number above images denote the number of feature maps in the layer and the dashed line shows which layer is concatenated in the up-sample part of network.}
 \label{fig:architecture}
 \end{figure*}

\subsection{Architecture} \label{Architecture}
A fully-convolutional network \citep[FCN,][]{2014Long} is a sequence of linear transformations of convolutional kernels with learnable weights and non-linear activation functions that retains the input size. In our work, we focus on a FCN known as U-net \citep{Ronneberger2015}. Our default architecture is the one from \cite{Chen2018}, where we have changed the number of input and output channels to correspond to our data.

As the name suggests, U-net has a U-shape created by down-sampling and up-sampling layers. The down-sampling part consists of four blocks of two convolution layers with each kernel size $3 \times 3$ followed by a Leaky Rectified Linear Unit (LeakyReLU) activation function  (described in \autoref{sec:act_function}). The next step is a $ 2 \times 2 $ max-pooling layer which down-samples the image size to half. After every down-sampling layer, the number of feature maps doubles. At the bottom of the network are two convolution layers followed by four up-sampling blocks. The up-sampling block is a $ 2 \times 2 $ transposed convolution layer whose output is concatenated with a sub-sample layer with the same shape and then two $3 \times 3$ convolutional layers with LeakyReLU activation. For the detailed breakdown of the architecture see \autoref{fig:architecture}. The last layer is a $1 \times 1$ convolution without an activation function. In \cite{Chen2018}, before the input images go through the network they are multiplied by the exposure time ratio between the input and ground truth image. In our architecture, we concatenate the feature map with matrix filled with values of the exposure time ratio into the first layer of transposed convolution (i.e. at the bottom of the U-net). For optimization, we choose the Adam optimizer \citep{Kingma2014}. The learning rate we use is $10^{-5}$ and after 2000 epochs it decreases to $10^{-6}$.  Each epoch has 150 iterations and at each iteration the network sees a random crop of size $256 \times 256$ from one training image.

The advantage of the FCN is that the input images can have different sizes, although large images require a lot of GPU memory. During evaluation, we use the same input size as that used during training.
In order to recreate the full image size we use a mosaic approach -- we take crop of the image with size $256 \times 256$, run it through the network, move across 32 pixels and take another crop etc. On overlapping areas we take the mean pixel value.

\subsection{Experiments} \label{subsec:experiment} 
During this project we trained more then 30 different networks with changes to the loss function, number of input/output channels, various exposure time ratios and different types of up-sampling.
In all the experiments we use an input size of $256 \times 256$ for computational efficiency. After every thousand epochs the PSNR and SSIM values are calculated and for comparison we choose the number of epochs with the highest score. The experiments in \autoref{loss} and \autoref{filters} are trained without information about the exposure time ratio and in \autoref{ETR} the networks are trained with exposure time ratio two. All networks were trained and evaluated for 5000 epochs.
\autoref{tab:star_detection} summarizes the results of the experiments done to find suitable hyper-parameters for the final networks. The results indicate that the change of the hyper-parameters can lead to a different performance of the network.

\begin{table*}
\caption{Comparison of different experimental networks and the performance metrics. During the experiments, the hyper-parameters of the network are optimised. }\label{tab:star_detection}
\centering
\begin{tabular}{cccccccccc}
\hline
Experiment & Description& RFE  [\%]&RFE uncertainty [\%] & TPR [\%]& F-measure &  $\mathrm{SNR_f}$ &PSNR [dB]&SSIM&KL \\ \hline
1 & $L1$ loss                 & 2.43 & 0.18 & 95.7 & 0.84 & 1.64 & 13.4 & 0.63 & 0.0074 \\
2 & $L2$ loss                 & 4.87 & 0.18 & 94.5 & 0.89 & 1.58 & 4.6  & 0.60 & 0.0130 \\
3 &$L1$ + perceptual loss     & 2.39 & 0.18 & 96.0 & 0.84 &	1.64 & 11.6 & 0.64 & 0.0069 \\
4 &$L1$ + KL divergence loss  & 2.18 & 0.18 & 95.9 & 0.85 &	1.63 & 14.0 & 0.64 & 0.0068 \\
5 &Filters                    & 2.66 & 0.18 & 96.3 & 0.87 &	1.62 & 11.6 & 0.64 & 0.0070  \\
6 & Input multiplied by $r$   & 2.39 & 0.18 & 96.0 & 0.84 & 1.62 & 12.7 & 0.64 & 0.0075 \\
7 & Segmentation map          & 8.52 & 0.16 & 95.0 & 0.91 & 1.58 & 12.4 & 0.61 & 0.0085 \\
8 &Multiple $r$ (random order)& 2.58 & 0.18 & 95.9 & 0.85 & 1.62 & 13.5 & 0.64 & 0.0071 \\
9 &ReLU activation            & 2.51 & 0.18 & 96.0 & 0.85 & 1.64 & 13.3 & 0.63 & 0.0072 \\
10 &PReLU activation          & 2.60 & 0.18 & 95.9 & 0.84 & 1.63 & 13.7 & 0.63 & 0.0070 \\
11 &SWISH activation          & 2.20 & 0.18 & 95.7 & 0.86 & 1.61 & 11.6 & 0.65 & 0.0073 \\\hline
\end{tabular}
\end{table*}

\subsubsection{Loss Function} \label{loss}
In order to find a suitable loss function, we test three options -- $L1$ loss (\textit{Experiment 1}), $L2$ (\textit{Experiment 2}) loss and perceptual loss \citep{Johnson2016} (\textit{Experiment 3}):

\begin{eqnarray}
L1 = \frac{1}{N} \sum_{i=1}^N \mid y_i-y'_i \mid,
\label{eq:l}
\end{eqnarray}

\begin{eqnarray}
L2 = \frac{1}{N}  \sum_{i=1}^N (y_i-y'_i)^2,
\end{eqnarray}

where the ground truth is denoted as $y$, the network output as $y'$ and $N$ is the number of pixels. The perceptual loss uses pre-trained networks for image classification. This loss function measures high-level perceptual and semantic differences between images. The architecture used in this test is similar to that described in \autoref{Architecture}, albeit without exposure time information. Our results show that the $L2$ loss is not suitable for our problem.  Moreover $L2$ loss does not correspond with characteristics of the human visual system. In \cite{Zhao2017} they show that $L2$ is more likely to get stuck in local minima whereas $L1$ reaches minima more efficiently. It is also important to note that the $L2$ loss is more sensitive to large errors. In cases where the distribution is likely to have outliers, it is generally more suitable to use the $L1$ loss as it is more robust.
In addition to all the hyper-parameters explored, we also combined the $L1$ loss function with the KL divergence (\textit{Experiment 4}).KL divergence is also sometimes called relative entropy, since it measures the gain (or loss) in information between two distributions. It is defined as:
\begin{eqnarray}
KL(p||q) =  \sum_{i=1}^N p_i \log\bigg(\frac{p_i}{q_i}\bigg),
\label{eq:kl_div}
\end{eqnarray}
where $p$ and $q$ are the true and predicted distributions respectively. If the distributions of the images being compared are the same, then the KL divergence will be zero. Minimizing the KL divergence is equivalent to the maximizing the likelihood between a Poisson model and the data.

\subsubsection{Filters} \label{filters}

The next test is dedicated to the number of input and output channels (\textit{Experiment 5}). Our data are captured with two different filters F555W or F606W. In our previous test, the input and output had one channel. Here, the output is still a one channel image, but now the input is a 2 channel image, where depending on if the filter used is F555W or F606W, the image will be placed in either the first or second channel respectively. The performance is similar to that of previous networks that have been tested; we do not find any gain in using multiple input filters.

\subsubsection{Exposure time ratio} \label{ETR}
We also wish to explore whether additional information of the exposure time ratio would help to generate better images, hence as previously mentioned the exposure time ratio is concatenated to the feature map at the bottom of the U-net. 
However, another way to include the exposure time ratio is to multiply it by the input image. In \textit{Experiment 6}, the input to the network is the image multiplied by the exposure time ratio. Alternatively, the exposure time ratio could also be added to the input image, however, we find that this delivers lower performance. Ultimately we choose to include the exposure time ratio concatenated into the bottom of the U-net, given the lower flux error that this option produces.

\subsubsection{Segmentation map}
In \textit{Experiment 7}, the input image has channels containing the image and the segmentation map, whilst the output still has only one channel containing the predicted image. The segmentation map is composed of objects separated from the sky background. The results show that including the segmentation map does not improve the results.

\subsubsection{Multiple exposure time ratios} \label{multilpe ETR} 
 We also investigate if it is possible to train the network with even higher exposure time ratios, and if so, what the most effective way to do this is. \textit{Experiment 8} uses the \textit{Network 1} setup, but the exposure time ratio is chosen in a \textit{random} order from ratio 2 to 5 during training. The evaluation is done with ratio 2 since all of the previous networks are trained with this ratio.
 Recall that in \textit{Network 2}, during each iteration, the input to the network consists of 4 identical crops with
exposure time ratios of 2, 3, 4, 5, which are fed into the network in that order. The results using ordered exposure time ratios are better than the network with randomly selected exposure time ratios, in terms of PSNR and flux error.

\subsubsection{Activation function} \label{sec:act_function}
The Leaky Rectified Linear Unit (LeakyReLU) is a commonly used activation function. We used this function as the base line for our project, but we wanted to test other activation functions as well. For all experiments in this section, we used the architecture of \textit{Network 1}. LeakyReLU is a variation on the ReLU activation function \citep{nair2010}. The advantages of ReLU over other previously proposed activation functions is that it is an easily optimized, monotonic function, unbounded above and bounded below. LeakyReLU is designed to avoid zero gradient, it leaks small negative numbers.  The function is define as:
\begin{eqnarray}
    \begin{split}f(x) = \begin{cases} x, & x > 0 \\
                     ax, & x <= 0 \end{cases}\end{split}
 \label{eq:relu}
\end{eqnarray}
where $a$ is zero for ReLU or a small constant number in the case of LeakyReLU, in our network we used $a=0.02$.

In \textit{Experiment 9} the ReLU activation is chosen.  Our results showed that LeakyReLU is a better option for our network, since networks using ReLU activations resulted in higher flux errors. 

In \textit{Experiment 10} we used ParametricReLU (PReLU) \citep{He2015}, for which parameter $a$ in \autoref{eq:relu} is a trainable parameter.
\textit{Experiment 11} used a self-gated activation function (SWISH, \cite{Ramachandran2017}). The authors show that using SWISH as an activation function matches or exceeds the results of ReLU on nearly all tasks. The functions are also unbounded above and bounded below, but SWISH is a non-monotonic function, which gives non-zero outputs for small negative inputs which improves gradient flow. It is defined as:
\begin{eqnarray}
    f(x) = x \cdot sigmoid(\beta \cdot x),
\end{eqnarray}
where $\beta$ is a constant or trainable parameter, in our experiment we use $\beta=1$.

\subsection{The best performing networks}  
In this section, we describe the two networks with the highest performance based on our experiments. The results of which, can be found in \autoref{tab:star_detection_result}. 

\subsubsection{Network 1} \label{sec:net1}
This is the baseline architecture that we have used throughout the experiments that was described in $\autoref{Architecture}$. The input and output have one channel with size $256 \times 256$. For training we use an exposure time ratio of two. The network is trained for 5000 epochs (\autoref{fig:loss}) and it takes $~48$ $hours$ on a \texttt{GeForce GTX 1080}. Each epoch has 160 iterations, where it sees a random crop from each of the 160 training images.  Validation of images are done after every thousand epochs and the number of epochs with the best results is chosen.

\subsubsection{Network 2}
\begin{figure}
\includegraphics[width=\columnwidth]{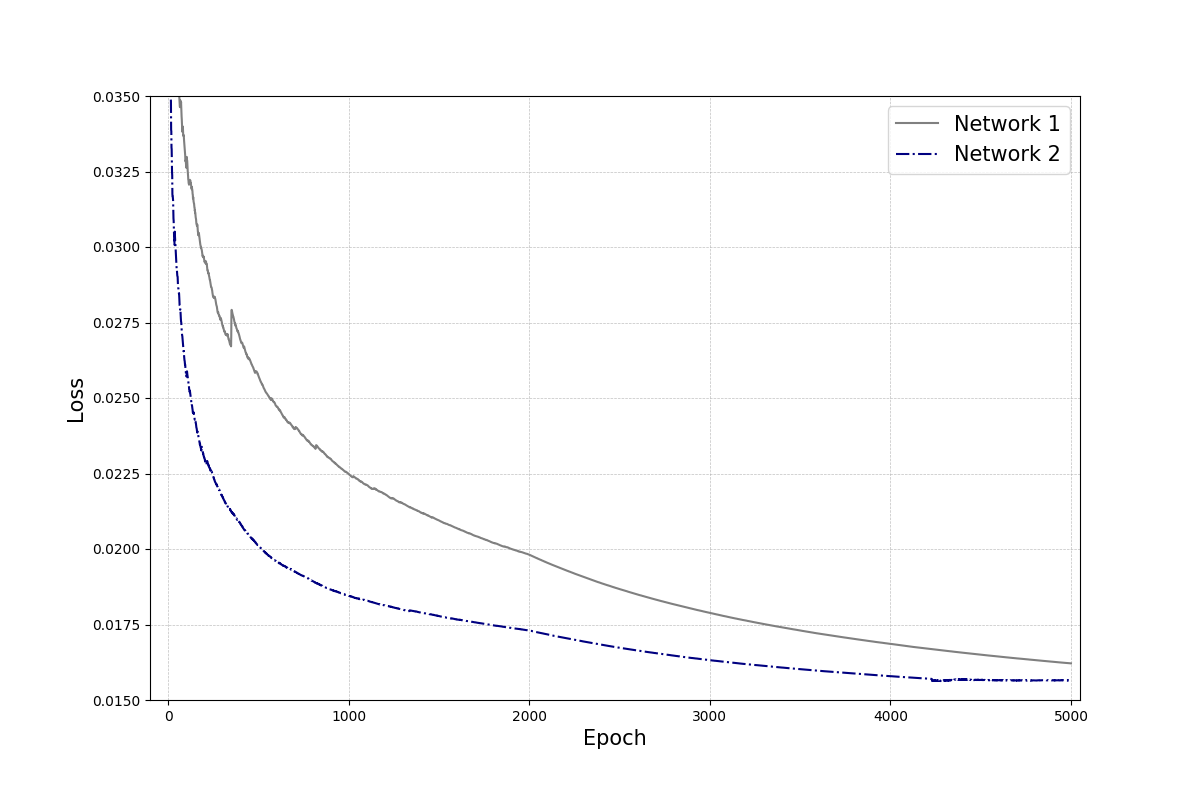}
\caption{Loss of \textit{Network 1} and \textit{2} using the $L1$ loss function (\autoref{eq:l}). \textit{Network 2} converges faster as a result of the network seeing more images in each iteration in comparison to \textit{Network 1}. The bump in loss curve of the \textit{Network 1} was caused by a technical problem that briefly paused the training.}
\label{fig:loss}
\end{figure}

In our second network, we use the same architecture but trained on exposure time ratios between two and five. The ratio is selected in order -- for every image crop we trained network with all ratios (2,3,4,5). \textit{Network 2} is trained for 5000 epochs (\autoref{fig:loss}), with 160 iterations per epoch. In each  iteration the network sees the same random image crop, with all exposure time ratios in order. Training of the network takes approximately 62 hours, again on a \texttt{GeForce GTX 1080}.

\section{Results} \label{Results}
This section contains description of several metrics used to evaluate the networks and the corresponding results.

\subsection{Metrics}
We use several metrics to evaluate the performance of our networks. The most common metrics for these methods are the Peak Signal-to-Noise Ratio (PSNR) and Structural SIMilarity Index \citep[SSIM]{Wang2004}.
PSNR measures quality of input or reconstructed image in comparison with the ground truth image, and is defined as:
\begin{eqnarray}
    MSE &=& \frac{1}{mn}\sum_{i = 0}^{m-1}\sum_{j = 0}^{n-1}[I(i,j) - I_\mathrm{Noise}(i,j)]^2,\\
    PSNR &=& 10 \log_{10}\bigg(\frac{MAX_I^2}{MSE}\bigg),
\end{eqnarray}
where $m\times n$ is the size of the image $I $ and the input image $ I_{Noise} $, and $MAX_I$ is the maximum value of image $I$. 
When the noise level of the ground truth image is high, our networks are able to recover images with even lower noise levels, and therefore PSNR can be misleading. For \texttt{RGB} images, PSNR cannot be negative, but since we are using \texttt{Fits} images that are not normalized to same pixel value range, the PSNR can be negative. SSIM takes into account the structure $s$ contrast $c$ and luminance $l$:
\begin{eqnarray}
    l(x,y) &=& \frac{2\mu_x\mu_y + c_1}{\mu_x^2 + \mu_y^2 + c_1},
\end{eqnarray}
\begin{eqnarray}
    c(x,y) &=&  \frac{2\sigma_{xy} + c_2}{\sigma_x^2+\sigma_y^2 + c_2},
\end{eqnarray}
\begin{eqnarray}
    s(x,y) &=& \frac{\sigma_{xy} + c_3}{\sigma_x\sigma_y + c_3},
\end{eqnarray}
\begin{eqnarray}
    c_1 &=& (k_1L)^2, 
\end{eqnarray}
\begin{eqnarray}
c_2 &=& (k_2L)^2, 
\end{eqnarray}
\begin{eqnarray}
    c_3 &=& \frac{c_2}{2},
\end{eqnarray}
\begin{eqnarray}
    SSIM(x,y) &=& [l(x,y)^\alpha \cdot c(x,y)^\beta \cdot s(x,y)^\gamma],
\end{eqnarray}
where $x$ and $y$ refers to the images of the same dimensions, $\mu$ is the mean and $\sigma$ is the variance with respect to their index, $L$ is the dynamic range of pixel values and $k_1$ and $k_2$ are small constants. $\alpha, \beta$ and $\gamma$ are weights and if we set them to 1, the equation can be reduced to:
\begin{eqnarray}
    SSIM(x,y) = \frac{(2\mu_x\mu_y + c_1)(2\sigma_{xy} + c_2)}{(\mu_x^2 + \mu_y^2 + c_1)(\sigma_x^2+\sigma_y^2 + c_2)}.
\end{eqnarray}

\begin{table*}
\caption{Results for \textit{Network} 1 and 2. The ratio denotes the exposure time ratio. We can see that the mean relative flux error RFE of \textit{Network 2} does not change significantly with different ratios, which indicates that the network learns to recover images with different scales of noise.}
\label{tab:star_detection_result}
\centering
\begin{tabular}{cccccccccc}
\hline
Images & Ratio & RFE [\%] & RFE uncertainty [\%] & TPR [\%]  & F-measure  & $\mathrm{SNR_f}$ & PSNR [dB] & SSIM & KL \\ \hline
Images with noise & 2 & 1.55 & 0.21 & 68.08 & 0.78 & 0.67  & $-16.0$ & 0.45   & 0.0231 \\
Network 1 & 2  & 2.26 & 0.18 & 95.94 & 0.86 & 1.63  & 13.6 & 0.64 & 0.0070\\
Network 2 & 2  & 2.26 & 0.18 & 96.26 & 0.85 & 1.64  & 16.3 & 0.63 & 0.0068\\
Network 2 & 3  & 3.40 & 0.18 & 94.27 & 0.84 & 1.94  & 15.0 & 0.53 & 0.0092\\
Network 2 & 4  & 3.78 & 0.18 & 94.21 & 0.80 & 2.28  & 14.6 & 0.49 & 0.0099\\
Network 2 & 5  & 4.13 & 0.18 & 93.26 & 0.78 & 2.55  & 14.4 & 0.46 & 0.0097\\ \hline
\end{tabular}
\end{table*}

\begin{figure*}
\includegraphics[width=\columnwidth]{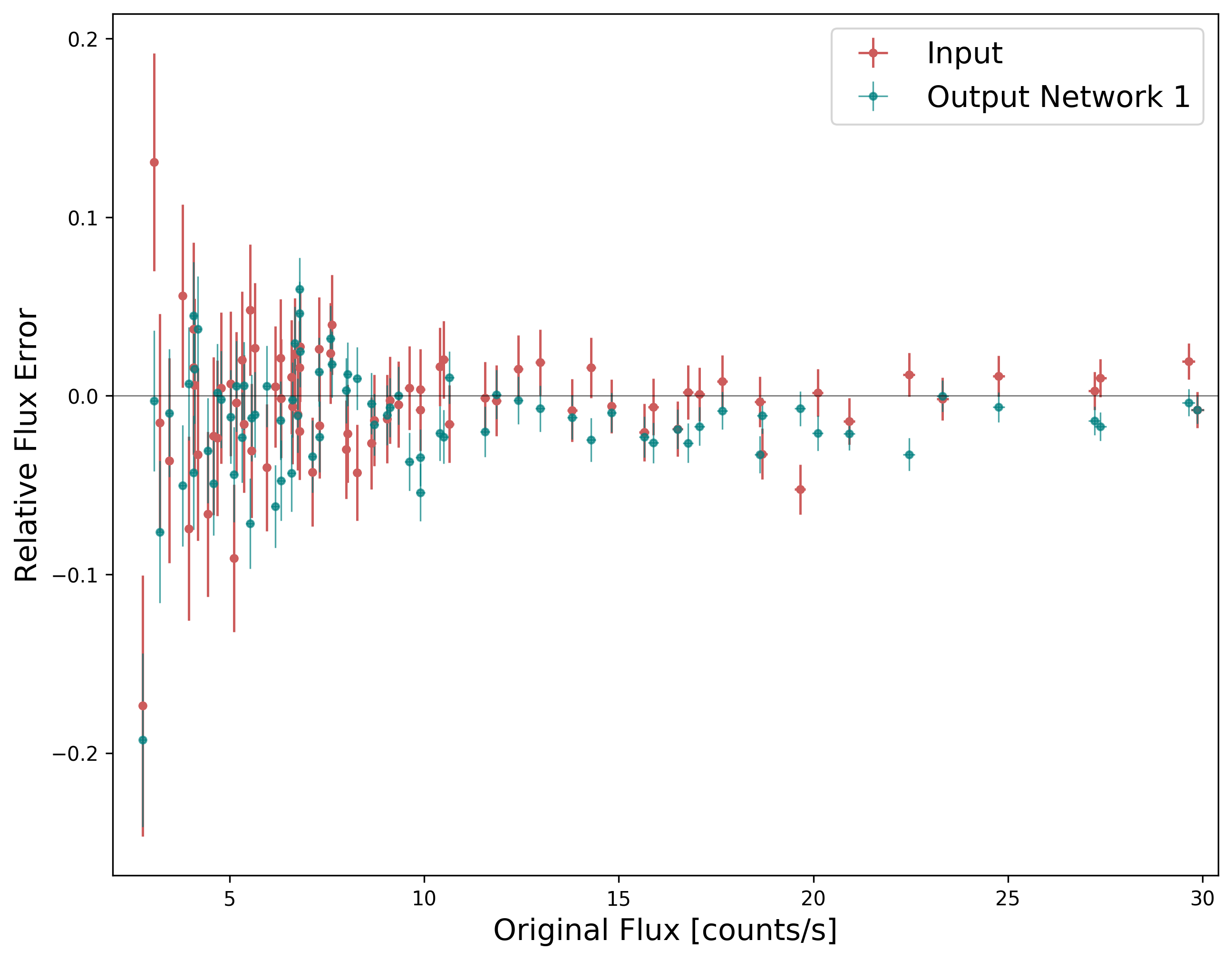}
\includegraphics[width=\columnwidth]{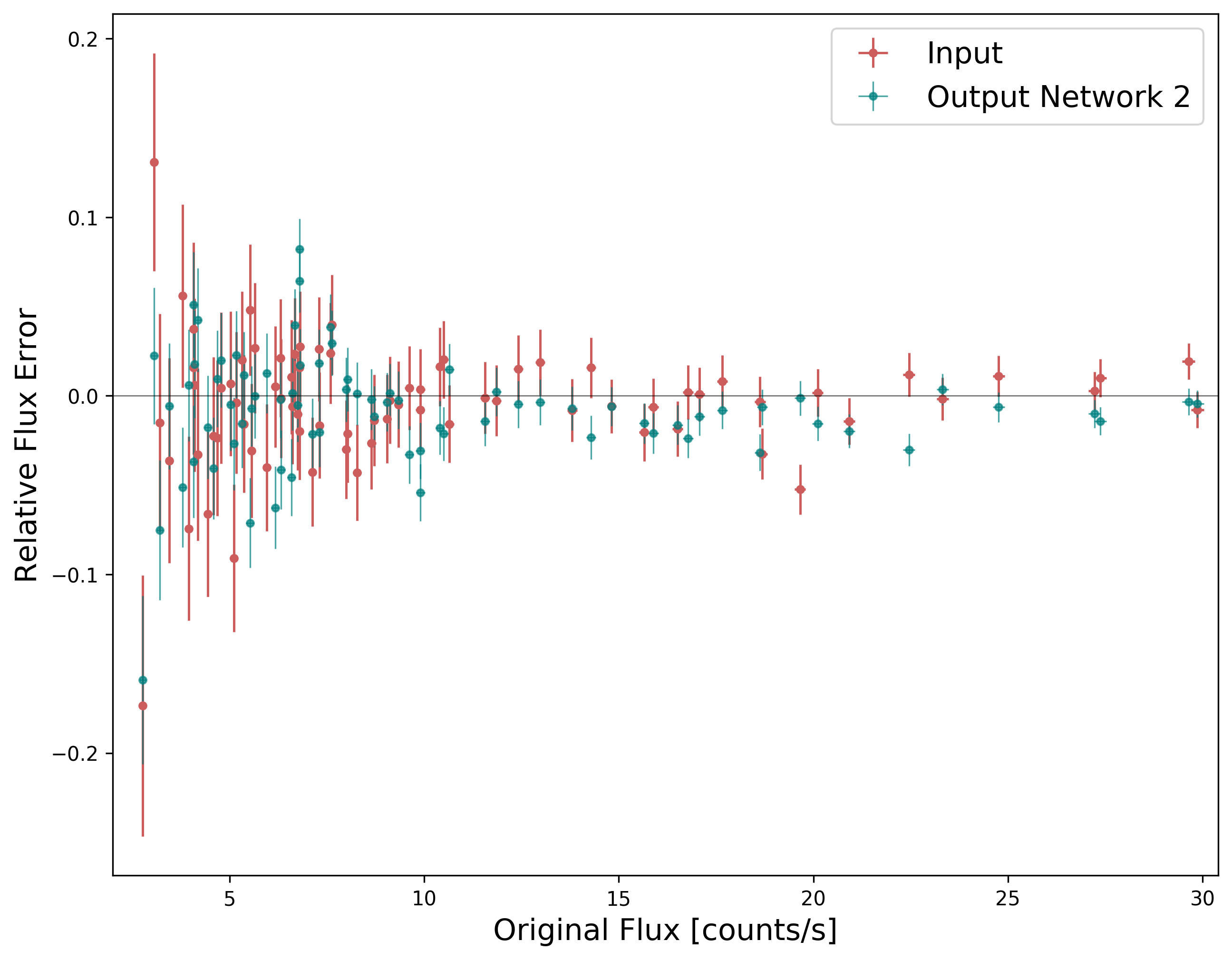}
\caption{Relative flux error versus star flux for stars cross-matched on the ground truth, output images and Input . From left: the y-axis is the absolute relative flux error and the x-axis is the flux of stars on the ground truth. The relative flux error (in percentage) for the input image is $1.39 \pm 0.19 \%$, for \textit{Network 1} $1.45 \pm 0.13 \%$ and \textit{Network 2 } $1.35 \pm 0.13 \%$}
\label{fig:flux_error_net}
 \end{figure*}

\subsection{Distribution}

\begin{figure}
\includegraphics[width=0.9\columnwidth]{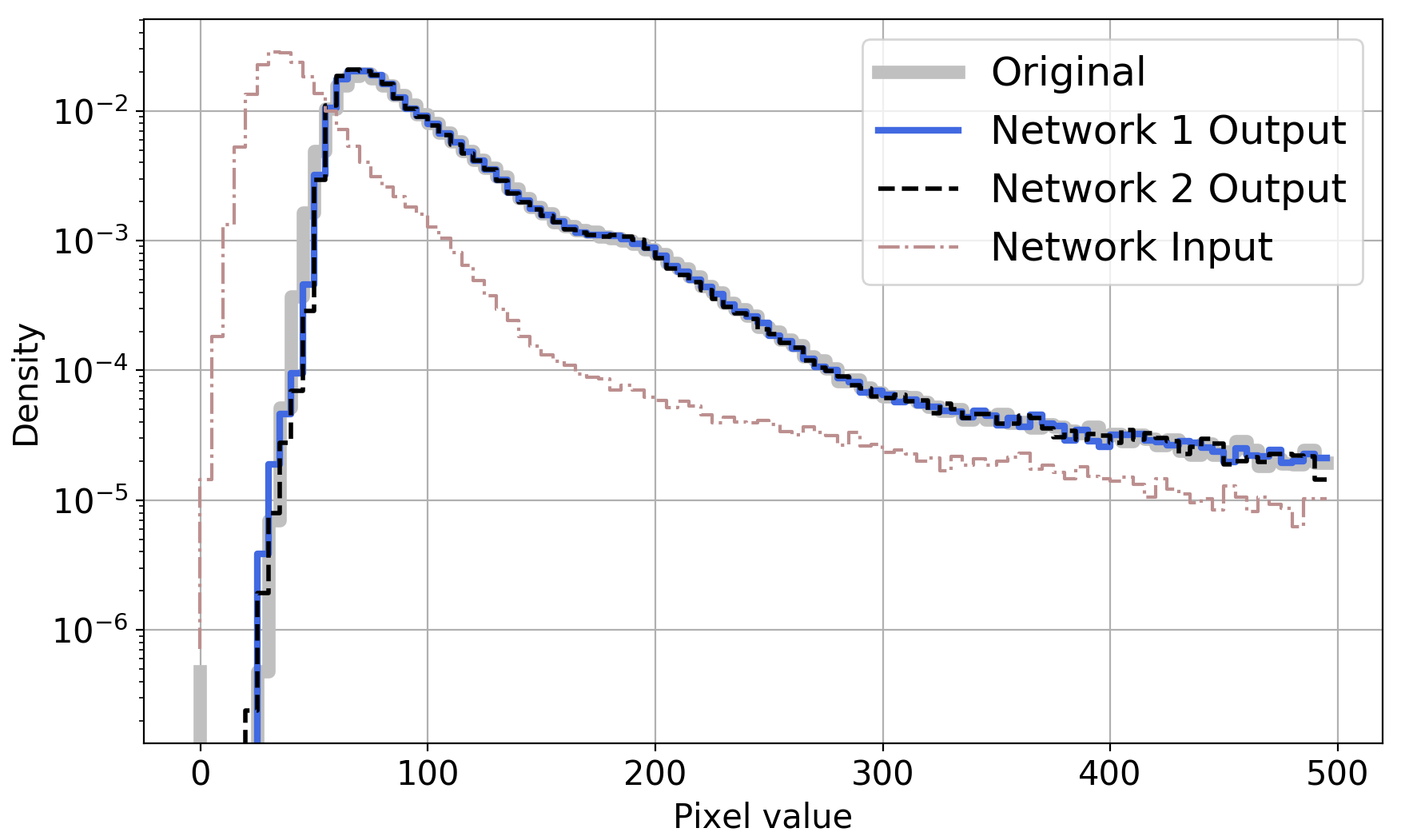}
\caption{Distribution of pixels in the ground truth image, the image with noise, and the images produced by \textit{Networks 1} and \textit{2}.}
\label{fig:histogram}
\end{figure}

We also examine the distribution of the image pixels by calculating the Kullback-Leibler (KL) divergence between the output and real image (\autoref{eq:kl_div}). 
We compare both the predicted image distribution and the noisy inputs to the ground truth image. We find that the KL divergence of the predicted images are lower. The output images of \textit{Network 1}  and \textit{Network 2}  have a KL divergence of$7 \times 10^{-3}$ and $6.8 \times 10^{-3} $ respectively and the KL divergence of the input is $2.3 \times 10^{-2}$. This indicates that the distribution of the network output images are closer to the image distribution of the ground truth than the input noisy image. An example pixel distribution is shown in \autoref{fig:histogram}.

\subsection{Source detection}
The most important part of the evaluation is the comparison of detected stars and their flux. We use \texttt{SExtractor} \citep{SExtractor1996} to detect stars and to compute their fluxes. We cross-match the detected stars between the images using \texttt{STILTS} \citep{Taylor2006}. True positives are defined as the stars in the true image that are also in the output (or input) image. False positives are stars that are detected in the the output (or input) image but are not found in the ground truth image. These stars are often not detected on the ground truth image (due to the noise settings of \texttt{SExtractor}) or are artefacts created by the network. False negatives are the number of stars in the true image that are not found in the output image. The fraction of true positives to the total number of stars in the true image is the true positive rate.
An alternative for true positive rate is the F-measure, which is a single score for precision and recall. It is often the preferred metric for an imbalanced dataset. Precision and recall are usually used in classification over accuracy, as the latter can be biased. They are defined as:
\begin{eqnarray}
Precision = \frac{True\ positives}{True\ positives + False\ positives},
\label{eq:precision}
\end{eqnarray}
\begin{eqnarray}
Recall = \frac{True\ positives}{True\ positives + False\ negatives},
\end{eqnarray}
the F-measure is then:
\begin{eqnarray}
F{\text -}measure = 2 \cdot \frac{Precision \cdot Recall}{Precision + Recall}.
\end{eqnarray}
Poor F-measure has a value close to zero, and in perfect performance takes the value of one.

For the true positives, we compare the flux of the detected stars in ground truth $F_{GT}$ and output $F_{O}$ image and compute the relative flux error $RFE$:
\begin{eqnarray}
RFE = \frac{F_\mathrm{GT}-F_\mathrm{O}}{F_\mathrm{GT}}.
\label{eq:error}
\end{eqnarray}
After we compute the relative flux error for the test images, we report the mean.

Another metric of the network performance is the measure of SNR of stars detected by \texttt{SExtractor}:
\begin{eqnarray}
SNR =\frac{1}{N} \sum_{i=1}^N\frac{F_i}{\sqrt{E_{i}^2+B^2}},
\label{eq:snr}
\end{eqnarray}
where $F_i$ and $E_i$ are the flux and the flux error of individual stars and $B$ is the background, as estimated by \texttt{SExtractor}. Here we use \texttt{FLUX\_APER} and the corresponding error to compute the SNR. The complete list of \texttt{SExtractor} parameters used are available on the github page\footnote{\url{https://github.com/Sponka/Astro_U-net}}. One way to improve SNR is to stack multiple images together, which improves SNR by a factor of $SNR_f$:
\begin{eqnarray}
SNR_\mathrm{f} = \sqrt{I} = \frac{SNR_\mathrm{stack}}{SNR_\mathrm{single}}, 
\label{eq:SNR_factor}
\end{eqnarray}
where $I$ is number of stacked images, $SNR_\mathrm{stack}$ and $SNR_\mathrm{single}$ denotes SNR of stacked image and SNR of single image respectively. We are curious how many input images are needed to be stacked together in order to obtain the same SNR as an image output by our network, so we modified \autoref{eq:SNR_factor}: 
\begin{eqnarray}
SNR_\mathrm{f} = \sqrt{I} = \frac{SNR_\mathrm{O}}{SNR_\mathrm{I}},
\label{eq:SNR_factor2}
\end{eqnarray}
here $SNR_\mathrm{O}$ refers to the SNR of the output image and $SNR_\mathrm{I}$ to the input image.  The results for \textit{Network 1} and \textit{2} are summarised in \autoref{tab:star_detection_result}.

\textit{Network 1} recovers $95.94\%$ of the detected stars in the true image and the mean stellar flux error is $2.26\%$. 
The true positive rate of stars detected on the real image are detected in the input image with half of the exposure time is $68.08\%$. These stars have relative flux error $1.55\%$. 
We find that the stars with the largest relative flux error are the dimmer ones that are not detected in the input images, and therefore the error on the input images and the network images seem to be similar however they are not directly comparable  (\autoref{fig:flux_error_net}). The mean SNR for the output image is $1.63$ times higher than the input image.  From \autoref{eq:SNR_factor}, we find that three input images are required to obtain the same SNR as our output image.

\begin{figure*}
\includegraphics[width=0.45\textwidth]{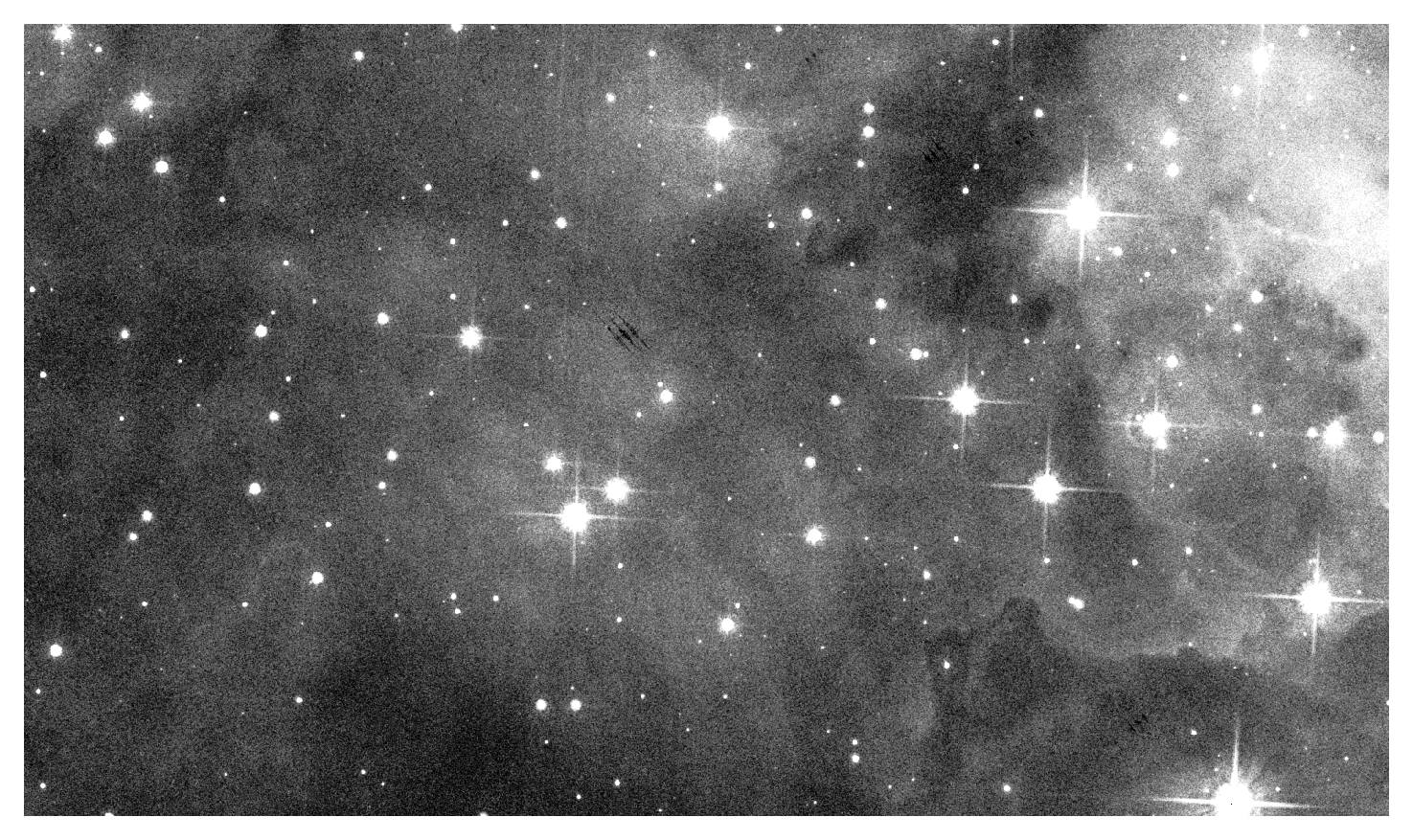}
 \includegraphics[width=0.45\textwidth]{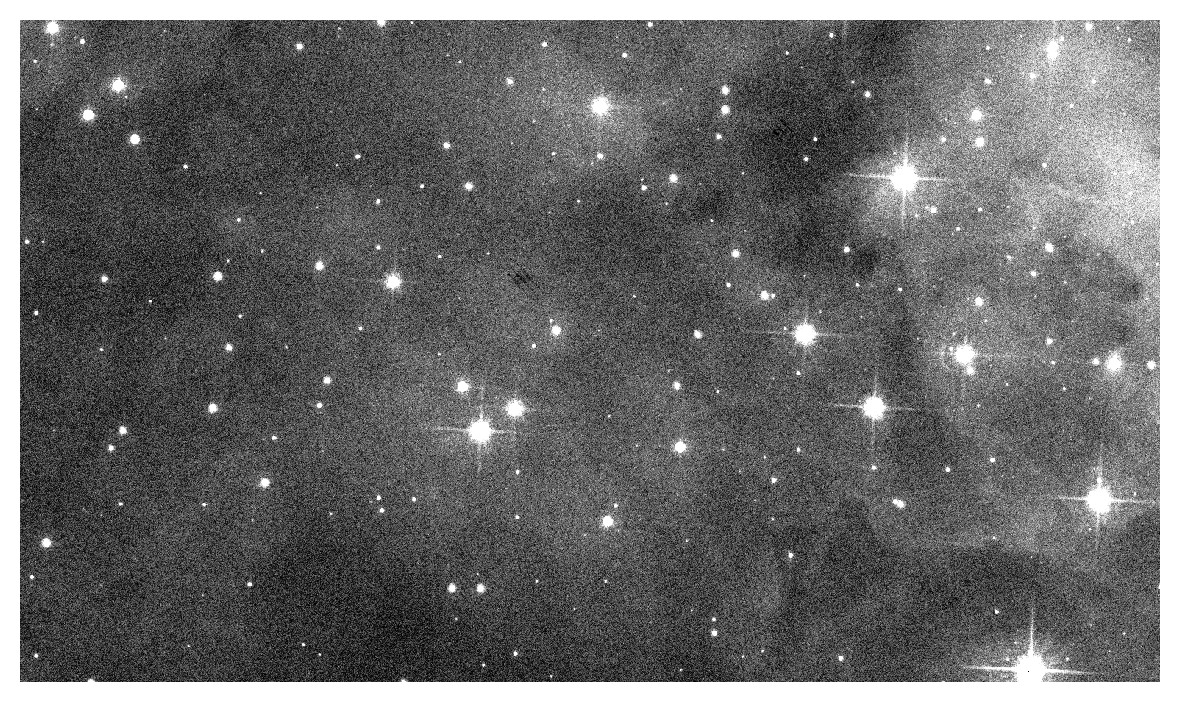}
 \includegraphics[width=0.45\textwidth]{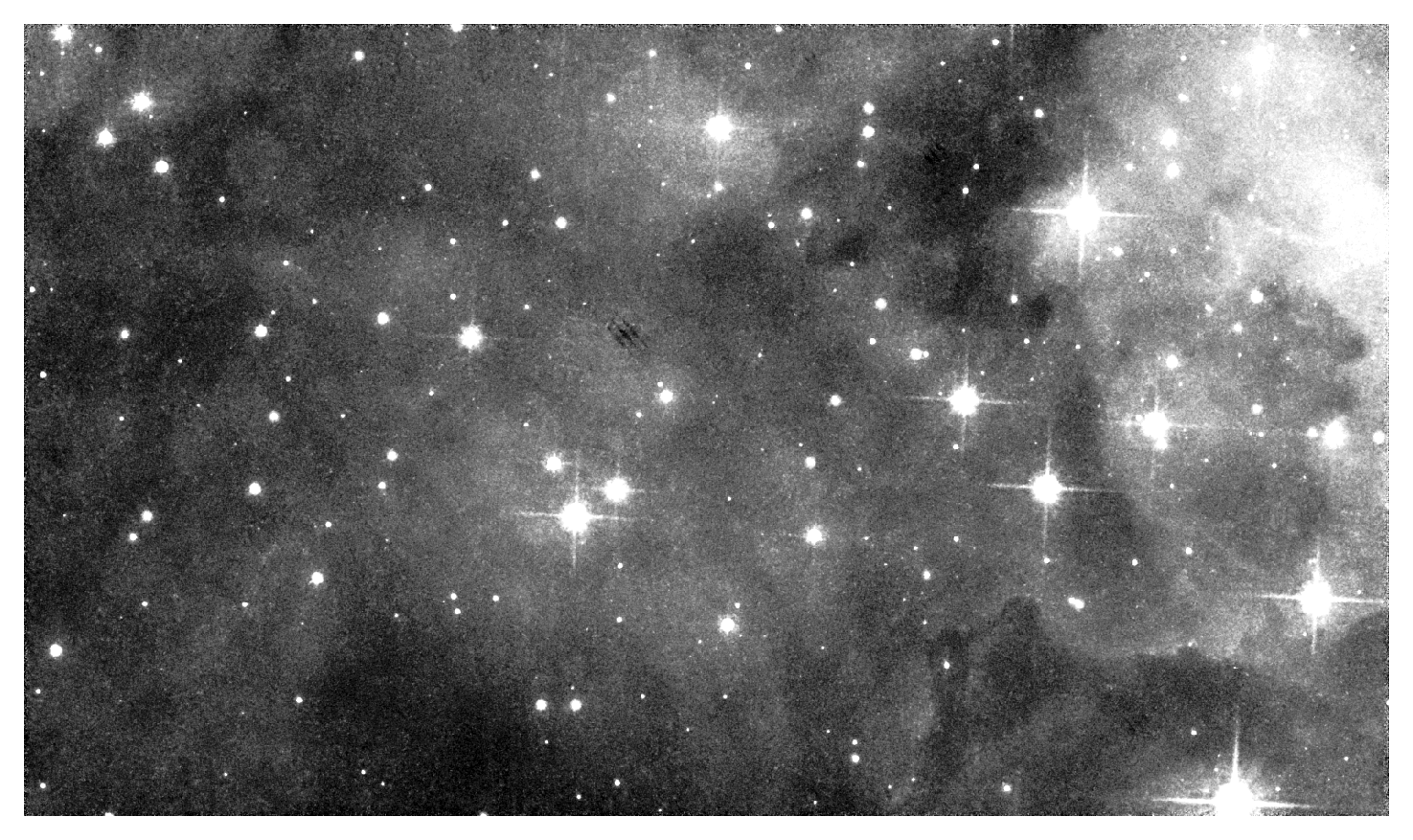}
 \includegraphics[width=0.45\textwidth]{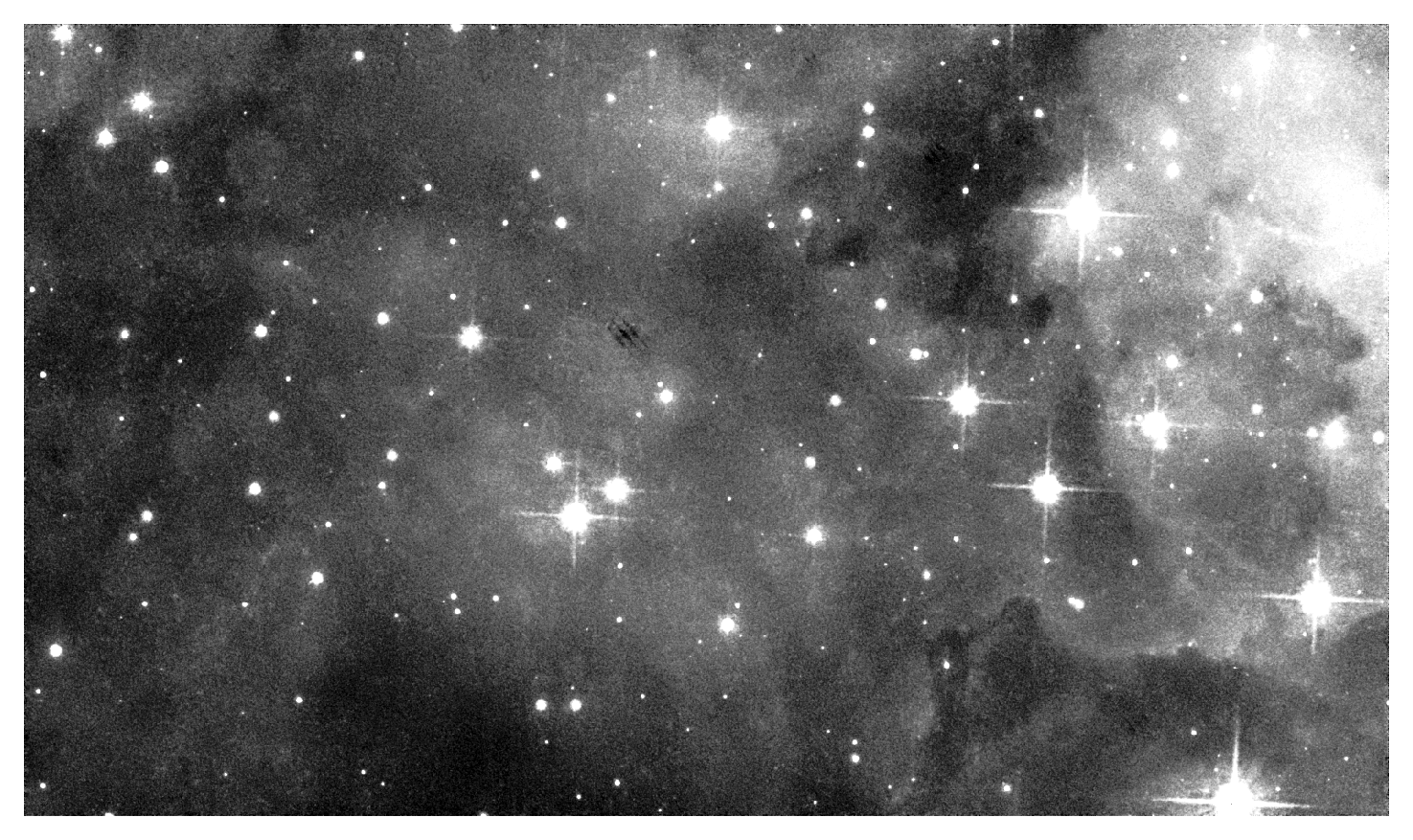}
 \includegraphics[width=0.45\textwidth]{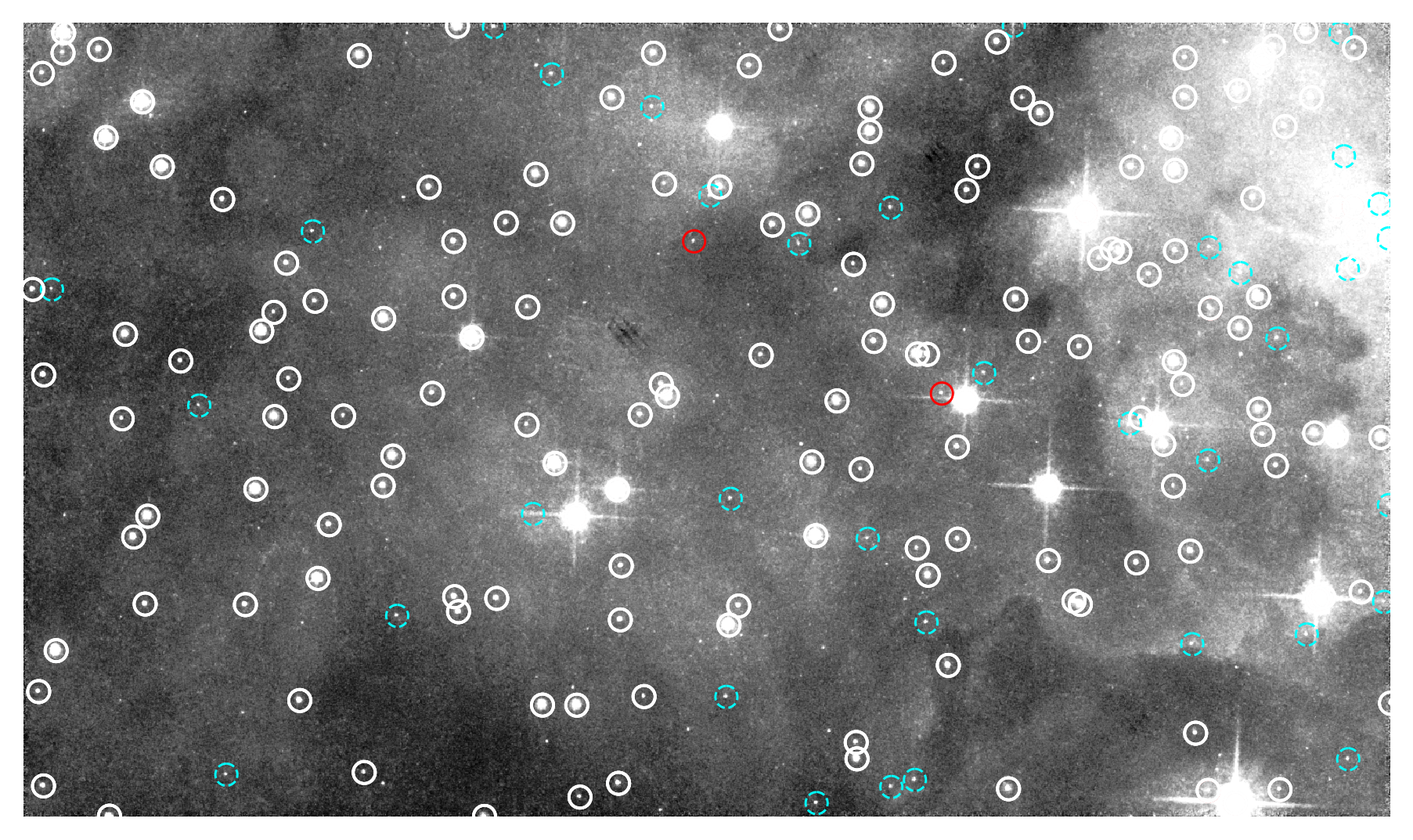} 
 \includegraphics[width=0.45\textwidth]{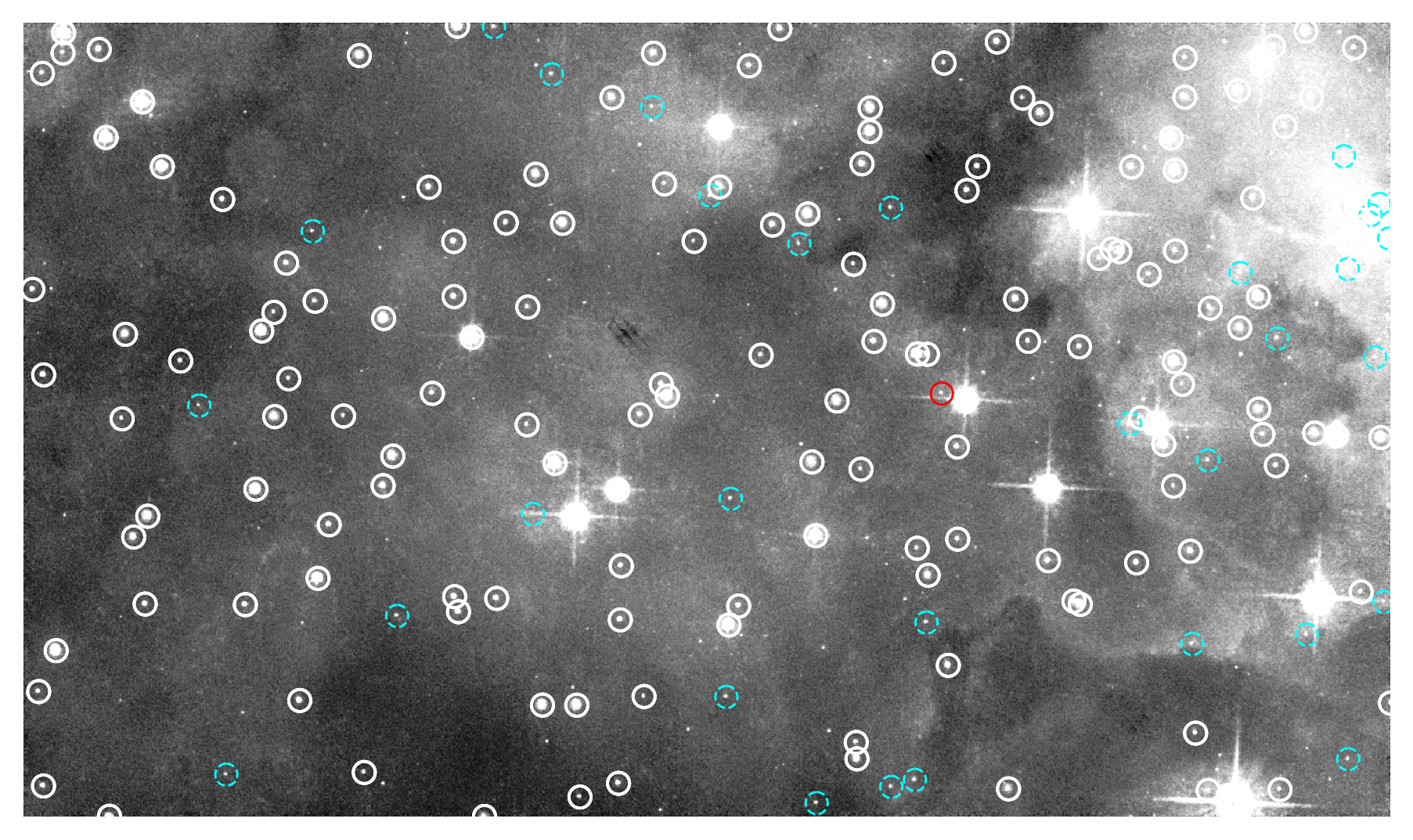}
 \caption{From the top left: the ground truth image and the input image with an exposure time ratio of two. Second row -- image reconstructed by  \textit{Network 1} and by \textit{Network 2}. The SNR for the images denoised by both networks is $\sim$ $1.5$ times higher than the SNR of the input image. Third row --  same image with stellar flux errors. The flux errors are $1.65\%$ and $1.70\%$ for \textit{Network} 1 and 2 respectively. The cyan circles enclose stars that are detected on the output image but not on the ground truth image, red enclose stars detected on ground truth but not on output image and the white circles enclose true positives. The image size is 36 $\times$ 61 arcsec. We encourage readers to look at the \texttt{FITS} files provided in the project github. }

 \label{fig:images}
 \end{figure*}

\begin{figure*}
 \includegraphics[width=0.45\textwidth]{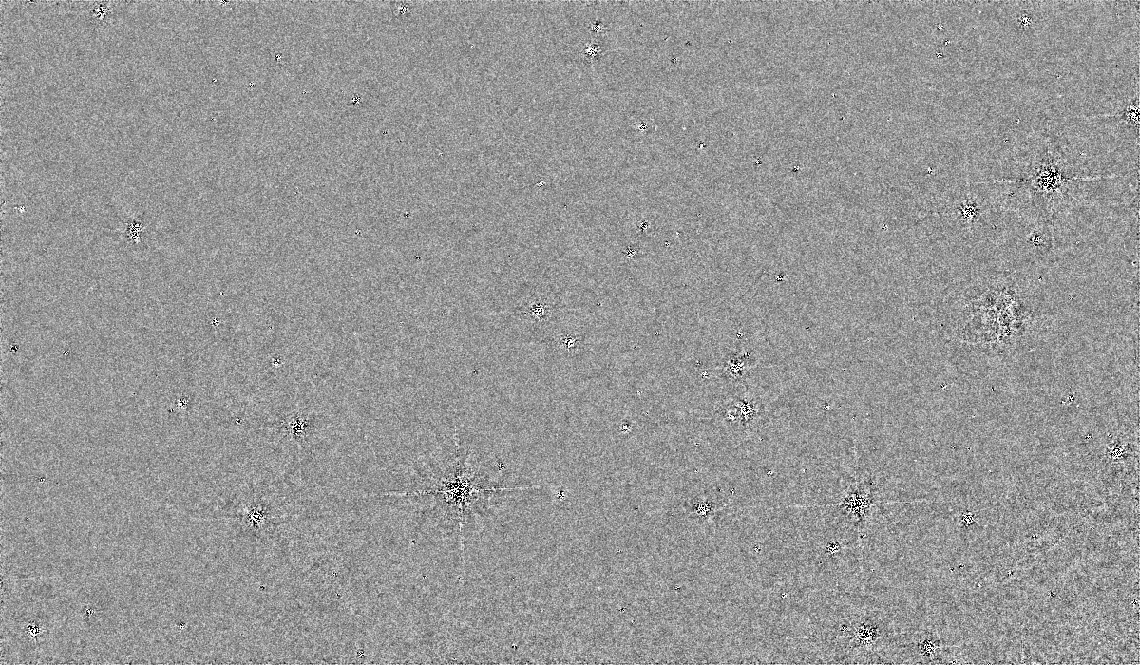}
 \includegraphics[width=0.45\textwidth]{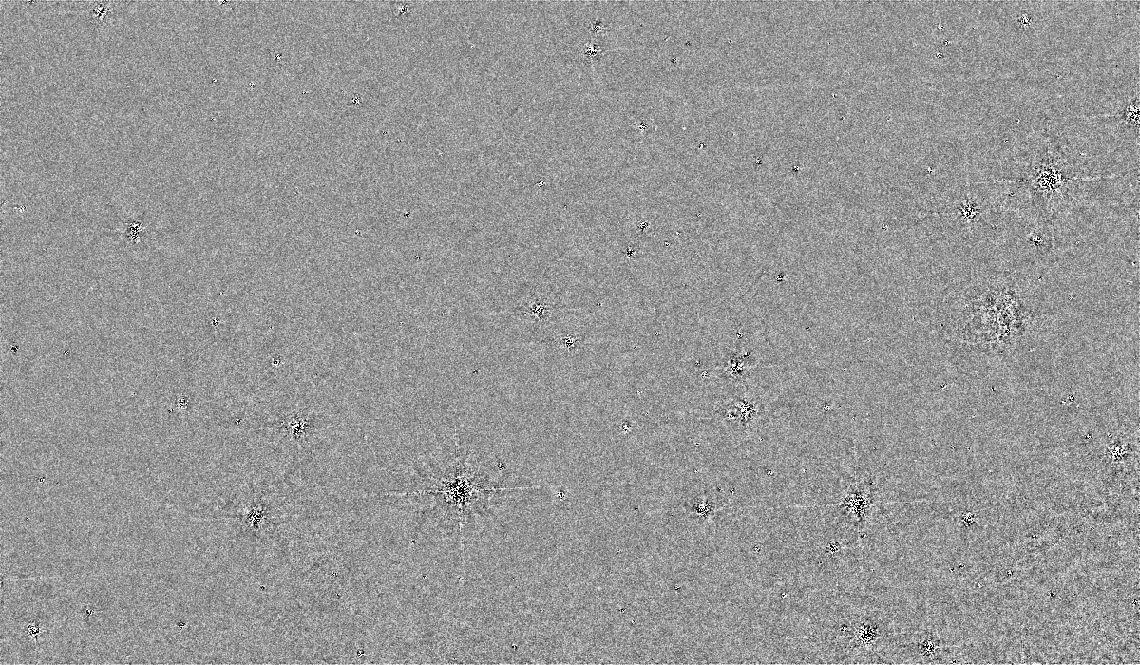}
 \caption{Residuals of \textit{Network 1} (left) and \textit{Network 2} (right). 
  Residual images are created as ground truth image minus the output image. The residuals tend to be higher in highly saturated areas e.g. stars, but there are no perceptible large-scale gradients. The mean values are 0.14, 0.26 and the standard deviation is 18.09, 13.32 for \textit{Network 1} and \textit{Network 2} respectively. Image size is 28 $\times$ 48 arcsec.}
 \label{fig:residuals}
 \end{figure*}

\textit{Network 2} has a true positive rate for ratio of two $96.26\%$ with a mean error of $2.26\%$. For an exposure time ratio of five, we achieve a $93.26\%$ true positive rate and a mean error of $4.13\%$ (\autoref{fig:flux_error_net}). With an exposure time ratio of two, the mean SNR of the output images is $1.64$ times higher than the SNR of input images, meaning on average, we would need to have almost $3$ input images to obtain an equivalent SNR to the output images. For an exposure time ratio of five, the SNR to the output images is $2.55$ times higher than the input images, which means that we would need at least 7 input images to obtain an equivalent SNR.
For both the ground truth and output images (both networks), the SNR $\sim$ 0.9, suggesting the information in the images are equal.
Examples of the reconstructed pixel distributions  and images are shown in \autoref{fig:histogram}  and \autoref{fig:images} respectively. The residuals (\autoref{fig:residuals}) suggests the networks have made good reconstructions, with poorer reconstructed areas corresponding to saturated pixels such as stars. 

\subsection{Real data}

\begin{figure*}
\includegraphics[width=0.4\textwidth]{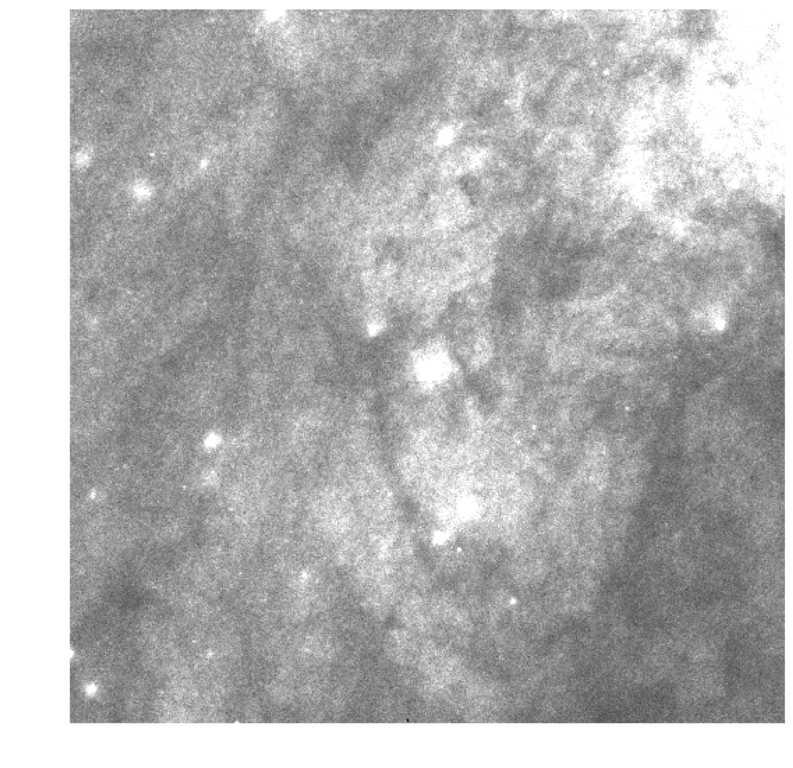}
 \includegraphics[width=0.4\textwidth]{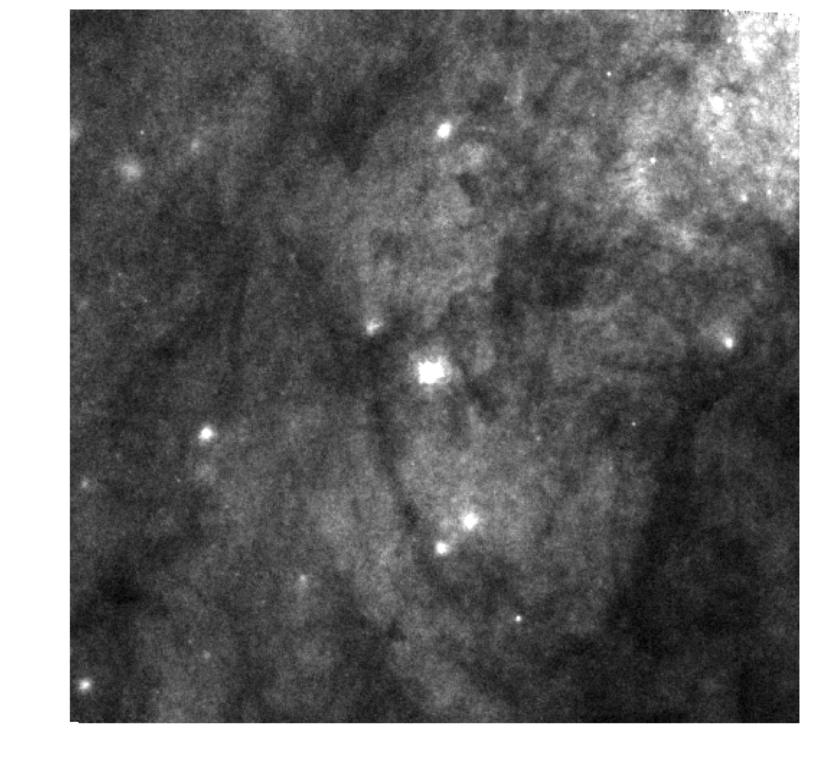}
 \includegraphics[width=0.4\textwidth]{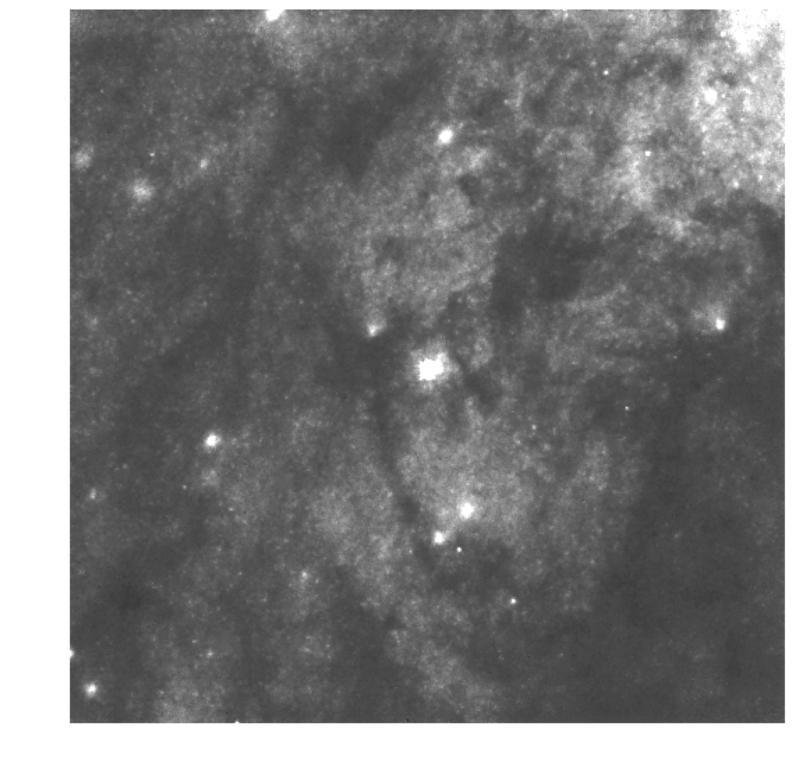}
 \includegraphics[width=0.4\textwidth]{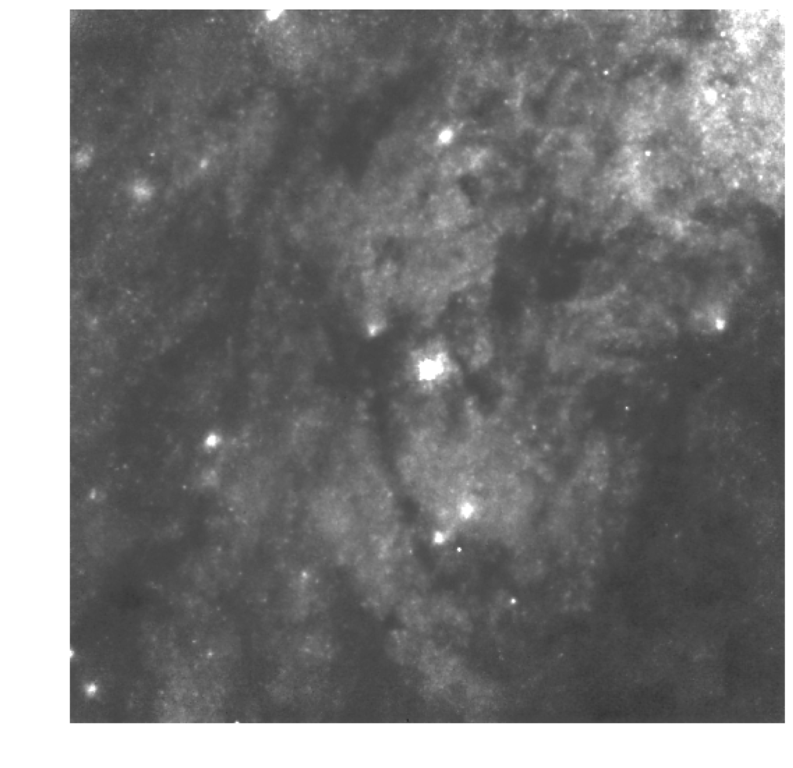}
 \caption{From top left: The real data image with exposure time $64s$, which is use as network input and $128s$ image for comparison. Second row -- images reconstructed by  \textit{Network 1} and by \textit{Network 2}. The size of the images is ~ 21 $\times$ 21 arcsec and the images are normalized for visualisation purposes, however we also make available the \texttt{Fits} files in the GitHub repository.}
 \label{fig:real_data}
\end{figure*}

\begin{figure*}
 \includegraphics[width=0.45\textwidth]{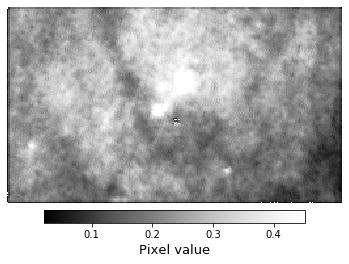}
 \includegraphics[width=0.45\textwidth]{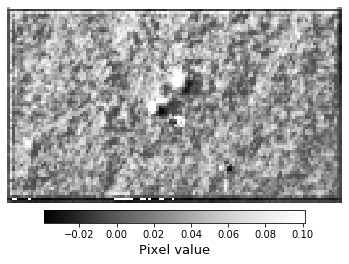}
 \includegraphics[width=0.45\textwidth]{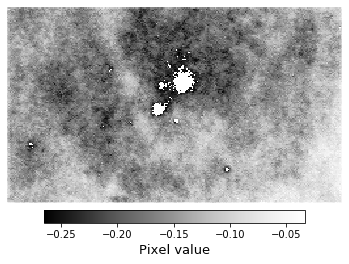}
 \includegraphics[width=0.45\textwidth]{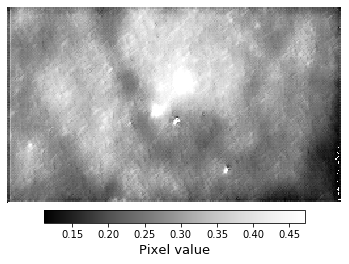}
 \caption{Examples of the feature maps from \textit{Network 1}. From the upper left, we show the feature map from the second and the fourth convolutional layer, below that, the second and third convolutional layers from the end of network. Here we have used the 64s image of SN2014J as input. The colour bars show the pixel values of the images. }
 \label{fig:conv}
 \end{figure*}

To demonstrate the power of our approach we also test our networks on real data, which they have not been trained on. We randomly select an observation of  SN2014J\footnote{\url{https://sky.esa.int/?target=148.9255\%2069.67386&hips=DSS2\%20color&fov=0.17480953085384782&cooframe=J2000&sci=true&lang=en}}, located in Messier 82 for this test. We use images taken with exposure times of $32, 64$ and $128$ seconds. Before we process the images, we use Astro-SCRAPPY \citep{2001Dokkum} to remove cosmic rays as the network tends to recover them as stars. \autoref{tab:real_image}  shows the results of a test where we use $32$ and $64$ second images as the input images for the network. For evaluation, we compare the network output with images with exposure times $64$ and $128$ seconds.
\autoref{fig:real_data} shows the images and \autoref{fig:conv} shows the feature maps of the convolutional layers of \textit{Network 1}. Although we have demonstrated that the network works on this particular image, the networks is trained on simulated data and so we cannot guarantee that it will work on all real data, we hope to develop this further in a future paper specifically for application to real data.

\subsubsection{Misaligned data}
The alignment of data is not a problem for synthetic data-sets, where the input and output are perfectly aligned. However in real world observations, there may be small misalignments between the input and ground truth. Pooling is known to help with shift invariance in images, none the less, we investigate the effects of small random misalignments of 0-4 pixels on the synthetic image pairs.  Using a KL-divergence + perceptual loss, we obtain a RFE of $~15\%$ and SNR $1.59$. Using smaller shifts (0-1 pixel) and L1 + KL divergence + perceptual loss, the RFE and SNR improves to $~3.6\%$ and $1.67$ respectively, and lastly with small shifts (0-1 pixel) and L1 + KL divergence loss but using transfer learning of the weights from \textit{Network 1} we obtain a RFE of $4.1\%$ and SNR of $1.67$. Visual comparison between output images created by these tests and ground truth images showed blurring which results in low PSNR, however the results are still promising.

\begin{table}
    \caption{The percentage of recovered stars and SNR of the images produced by Network 1 and 2 based on input images of different exposure times. For images assigned as Network 1 or 2, the Exp time column is the exposure time of the input image, and for images assigned as Input this column it is their exposure time.}
    \label{tab:real_image}
    \centering
    \begin{tabular}{ccccc}
    \hline

         Image         & Exp time [sec] & RFE [\%]  &  TPR [\%]    & $\mathrm{SNR_f}$ \\\hline
         Input    & $32$        & 2.4         & 100           & 0.75 \\
         Input    & $64$        & 5.2         & 56.2          & 0.65 \\
         Network 1& $32$        & 3.0         & 100           & 1.47 \\
         Network 1& $64$        & 9.8         & 87.5          & 1.31 \\
         Network 2& $32$        & 2.7         & 100           & 1.53 \\
         Network 2& $64$        & 9.8         & 87.5          & 1.34 \\ \hline
    \end{tabular}
\end{table}

\section{Discussion} \label{Discussion and conclusion}

\subsection{Similar work}
We are not the first to attempt to denoise astronomical data. Using a a 3-layer CNN with ReLU activation function, \cite{Flamary2016} was able to denoise $32 \times 32$ images albeit to a lower resolution output ($14 \times 14$). Their network has less layers then ours, and therefore is faster to train and test but still obtaining a good PSNR and outperforming traditional approaches such as the Wiener filter \citep{Starck_2002}, Richardson-Lucy algorithm \citep{Richardson:72, Lucy_1974} and Total Variation regularization \citep{condat}. \\
\cite{Schawinski2017} introduced Generative Adversarial Networks \citep[GANs, ][]{Goodfellow2014} for recovering features in astrophysical images, where a generator network creates images and a discriminator network serves as a loss function. The architecture of the generator is more similar to ours in that they use transposed convolution and the output image is able to retain the input image resolution. This method too was shown to outperform the Richardson-Lucy algorithm and the Blind Deconvolution \citep{bell1995} in feature recovery however, the network is not able to handle \texttt{FITS} images.\\ 
Our network both denoises and enhances an image to obtain images emulating longer exposure times, and therefore we are unable to make a direct comparison to these classical approaches. A fair comparison of the performance would require the adoption of the  \cite{Flamary2016} and \cite{Schawinski2017} architectures, however this is beyond the scope of this work.

\subsection{Conclusion and Future work}
In this article we present a new approach to de-noising and enhancing astronomical images with the aim of reducing the telescope exposure times needed for science analysis. In comparison with current methods that use stacking of large numbers of exposures, our method, \texttt{Astro U-net}, is less time consuming and is able to yield results of equivalent quality. \texttt{Astro U-net} is a fully-convolutional neural network for image de-noising and enhancement, which only requires the exposure time ratio as input. Moreover, \texttt{Astro U-net} can handle images of different scales.

Never the less, \texttt{Astro U-net} does have its limitations. The network is trained on the HST images from the WFC3 instrument UVIS with F555W and F606W filters. To adopt the network to a different data set would require retraining of the network on the new data, or the use of transfer learning \citep{2010Pan} to fine-tune parameters. Another problem can be caused by cosmic rays, which can be recovered by the network as stars. We encourage the removal of cosmic rays before using the network. Due to the presence of artefacts near to the edges of our training images, our output images can also incur some artefacts, However despite these limitations we demonstrate that \texttt{Astro U-net} can be used for science applications. 

Our experiments show that there are still many opportunities to improve upon our pipeline and obtain better results. In the near future we plan to create a data set from real astronomical data and use transfer learning \citep{2010Pan} to train the network. Additionally we want to apply machine learning for super resolution of the astronomical images \citep{2019Zhang}.

\section*{Acknowledgements}
AV acknowledges a ESA traineeship at ESAC. ML and LO acknowledge a ESA research fellowship at ESAC and ML, a research fellowship at the University of Nottingham. 

\section*{Data availibility}
The datasets were derived from sources in the public domain. This data and the source code are available in the article and in its online supplementary material. 



\bibliographystyle{mnras}
\bibliography{main} 


\bsp	
\label{lastpage}
\end{document}